\documentclass[twocolumn,aps,prr,groupedaddress,superscriptaddress,amsmath,amssymb,showpacs,noeprint,english]{revtex4-2}
\usepackage[T1]{fontenc}
\usepackage[latin9]{inputenc}
\usepackage{amsmath}
\usepackage{amssymb}
\usepackage{mathrsfs}
\usepackage[dvipsnames]{xcolor}
\usepackage{ulem}
\usepackage{esint}
\usepackage{soul}
\usepackage{color}

\newcommand{\ome}{\omega_r}
\newcommand{\para}{\eta}

\makeatletter
\usepackage{amsmath}    
\usepackage{graphicx}   

\makeatother

\usepackage{babel}

\begin{document}

\title{Protecting operations on qudits from noise by continuous dynamical
decoupling}

\author{Reginaldo de Jesus Napolitano}

\affiliation{S\~{a}o Carlos Institute of Physics, University of S\~{a}o Paulo, PO Box
369, 13560-970, S\~{a}o Carlos, SP, Brazil}

\author{Felipe Fernandes Fanchini}
\email{felipe.fanchini@unesp.br}

\affiliation{Faculdade de Ci\^{e}ncias, UNESP - Universidade Estadual Paulista, 17033-360
Bauru, S\~{a}o Paulo, Brazil}

\author{Adonai Hilario da Silva}

\affiliation{S\~{a}o Carlos Institute of Physics, University of S\~{a}o Paulo, PO Box
369, 13560-970, S\~{a}o Carlos, SP, Brazil}

\author{Bruno Bellomo}
\affiliation{Institut UTINAM, CNRS UMR 6213, Universit\'{e} Bourgogne Franche-Comt\'{e}, Observatoire des Sciences de l'Univers THETA, 41 bis avenue de l'Observatoire, F-25010 Besan\c{c}on, France}

\begin{abstract}
We develop a procedure of generalized continuous dynamical decoupling (GCDD) for an ensemble of  $d$-level systems (qudits), allowing one to protect the action of an arbitrary multi-qudit gate from general noise. We first present our GCDD procedure for the case of an arbitrary qudit and apply it to the case of a Hadamard gate acting on a qutrit. This is done using  a
model that, in principle, could be implemented using the three magnetic hyperfine states of the ground energy
level of $^{87}\mathrm{Rb}$ and laser beams whose intensities and
phases are modulated according to our prescription. 
We show that this model allows one to generate continuously all the possible SU(3) group operations which are, in general, needed to apply the GCDD procedure.	
We finally show that our method can be extended to the case of
an ensemble of qudits, identical or not.
\end{abstract}

\maketitle
\section{Introduction}
The development and implementation of effective quantum computers
is of great interest to the scientific community as well as to the
world economy as a whole \cite{Acin2018}. Indeed, quantum computers promise to revolutionize
many important tasks, even with a reduced number of algorithms known
to be more efficient than their classical analogues \cite{BB01}.
In this sense, reducing errors while keeping quantum algorithms simple
is an important aspect to be addressed. Several strategies have been proposed to contrast decoherence effects, such as  reservoir engineering methods \cite{Plenio2002, Braun2002,Bellomo2008,Verstraete2009,Krauter2011,Bellomo2015}, optimal quantum control protocols \cite{Reich2015,Glaser2015}, measurement-based control \cite{Yan2013,Wang2014,Wakamura2017}, and pulsed dynamical decoupling of qubits \cite{Viola1998,Viola1999a,Viola1999b,Viola2004}.
 Some of us worked at continuous dynamical decoupling strategies, applied to single- and two-qubit systems \cite{Fanchini2007a,Fanchini2007b,Fanchini2007c,Fanchini2015}.

Continuous dynamical decoupling techniques have been theoretically investigated and experimentally implemented in several contexts. For example, these kinds of techniques have been applied in the case of nitrogen-vacancy centres  to separate a single nuclear spin signal from the bath noise \cite{exp1}, to extend the coherence time for the electron spin \cite{exp2,exp3} while protecting quantum gates \cite{exp3}, to provide single-molecule magnetic resonance spectroscopy \cite{exp4}, and in the context of sensing high frequency fields \cite{exp5}. Furthermore, they can be used to create a dephasing-insensitive quantum computation scheme in an all-to-all connected superconducting circuit \cite{exp6}, and for engineering an optical clock transition in trapped ions, robust against external field fluctuations \cite{exp7}.

Although qubits
are virtually ubiquitous in the
development of quantum algorithms \cite{BB01}, $d$-level systems
(qudits) seem to be potentially more powerful for information processing
\cite{BB02,BB03,BB04,BB05,BB06,BB07,BB08,BB09,BB10,BB11,BB12,BB13,BB14}.
Indeed, the use of higher-dimensional quantum systems brings significant
advantages, allowing for information coding with increased density
and thus reducing the number of multi-particle interactions. Specifically,
the use of qudits brings improvements in the building of quantum logic
gates and the simplification of the design of circuits \cite{BB02,BB03,BB04,BB05},
in the security of quantum key distribution protocols \cite{BB06,BB07,BB08,BB09,BB10},
in performing quantum computation \cite{BB11,BB12,BB13,BB14,BB15},
as well as in the realization of fundamental tests of quantum mechanics
\cite{BB16}. In particular, powerful error correction procedures
have been proposed for qudits \cite{BB17,BB18,BB19,BB20}. We remark
that some of the above advantages have been pointed out already for
the case of  qutrits (see, e.g., \cite{BB03,BB04,BB06,BB07}).
Several setups have been considered to experimentally implement
qudits, including optical systems \cite{BB21,BB22,BB23}, superconductors
\cite{BB05,BB24}, and atomic spins \cite{BB25,BB26}.

In this paper, we present a complete theoretical prescription for a generalized
continuous dynamical decoupling (GCDD) of an ensemble of arbitrary qudits from environmental noise in the presence of  an arbitrary quantum operation. 
 Our prescription consists of a continuously-varying
control Hamiltonian and a modification of the intended quantum operation.
We first develop our procedure for the case of an arbitrary qudit and apply it to the case of a qutrit implemented with a specific model where we explicitly show how to realize  all the possible SU(3) group operations which are, in general, needed for our scheme. We  finally show that our scheme can be extended to the case of an ensemble of qudits, identical or not. Since our procedure works for an arbitrary number of levels and can be generalized to the case of many qudits, it extends the range of applicability of continuous dynamical decoupling strategies to more complex systems with respect to previous studies.

The paper is organized as follows. In Sec.~\ref{Sec: The GCDD procedure} we present the GCDD procedure for the specific case of a qudit. In Sec.~\ref{Sec: Our prescription} we give our prescription to build up the control fields necessary for the GCDD procedure. In Sec.~\ref{Sec: atomic qutrit} we show how to apply our GCDD procedure in the case of an atomic qutrit realized with some states of $^{87}\mathrm{Rb}$. In Sec.~\ref{numsim} we present our numerical simulations showing the protection realized by our GCDD procedure in the case of the atomic qutrit previously presented, in the presence of a noise taking into account both damping and dephasing effects. In Sec.~\ref{Sec: Extension to the case of a multi-qudit scenario} we comment on the possibility to extend our GCDD to the case of an ensemble of many qudits, identical or not, by referring to  Appendix~\ref{SeveralQudits} where the case of two qudits is explicitly treated. Some parts of our analysis are reported in various Appendices. 

\section{The GCDD procedure}\label{Sec: The GCDD procedure}

Here, we present our GCDD procedure for the specific case of an arbitrary qudit of dimension $d$.

Let $H_{G}$ be the Hamiltonian generating the intended
evolution of an arbitrary qudit state, in the
ideal, noise-free case. Notice that the free Hamiltonian of the qudit can be included in $H_G$ or assumed to be eliminated before the GCDD procedure. The procedure is then directly applied to a  system which, in the absence of $H_G$, is degenerate since all qudit levels have the same energy. After a gate operation time $\tau$,
the desired evolution operator acting on an initial qudit state is then
given by $U_{G}=\exp(-iH_{G}\tau/\hbar)$. Instead of $H_{G}$, we aim to use external fields whose
interaction with the qudit is described by a non-autonomous Hamiltonian
$H_{\mathrm{lab}}(t)$, acting continuously during the time interval
$\tau$ and only on the qudit Hilbert space, and generating, despite the presence
of noise, an effective evolution of the qudit that, at
least up to a high enough fidelity, is at time $\tau$ the same as the one  provided by $U_{G}$.

To describe $H_{\mathrm{lab}}(t)$, we first split this control field
Hamiltonian into two terms, $H_{\mathrm{lab}}(t)\equiv H_{c}(t)+H_{\mathrm{gate}}(t)$, where $H_{c}(t)$ is the control Hamiltonian that will continuously decouple the qudit evolution from the interference of the environment, while $H_{\mathrm{gate}}(t)$ will
provide the modified gate Hamiltonian that in the end will effectively
reproduce the action generated by $H_{G}$. Associated
with $H_{c}(t)$ there is a unitary operator $U_{c}(t)$
that we require to be periodic with a period $t_{0}$ and that should  satisfy (but non necessarily)
the dynamical-decoupling condition \cite{Facchi}
\begin{equation}
\int_{0}^{t_{0}}dt\,\left[U_{c}^{\dagger}\left(t\right)\otimes\mathbb{I}_{E}\right]H_{\mathrm{int}}\left[U_{c}\left(t\right)\otimes\mathbb{I}_{E}\right] =  0,\label{eq:dd}
\end{equation}
where $\mathbb{I}_{E}$ is the identity operator of the environmental
Hilbert space and $H_{\mathrm{int}}$ is the Hamiltonian interaction term
coupling the qudit with its environment. It is important to emphasize that a weaker, but sufficient, condition would also work to obtain dynamical decoupling. 
This weaker condition can be written as 
\begin{equation}
\int_{0}^{t_{0}}dt\left[U_{c}^{\dagger}\left(t\right)\otimes\mathbb{I}_{E}\right]H_{\mathrm{int}}\left[U_{c}\left(t\right)\otimes\mathbb{I}_{E}\right] = c \, t_{0} \,\mathbb{I}_{d}\otimes B,\label{conditionweak}
\end{equation}
where $\mathbb{I}_{d}$ is the identity operator of the qudit Hilbert space of finite dimension $d$, $c$ is a constant, and $B$ is some operator that acts on the environment.
Indeed, if the integral of Eq.~\eqref{eq:dd} results  to be proportional to a tensor product between the qudit identity and an environment operator, the system will also be protected.
To explicitly illustrate this aspect, we consider in detail this situation in Appendix \ref{Sec:Dynamical decoupling condition}, where we show that also in this case the qudit dynamics is protected from the effects of the coupling with the environment. In the following we show that our protection scheme satisfies this weaker condition. We finally observe that we choose $t_{0}$ such that $\tau$ results to be an integer multiple of $t_{0}$ itself, for reasons that we explain below.

The total Hamiltonian of the system and the environment is then written as
\begin{equation}
H_{\mathrm{tot}}\left(t\right)  =  \left[H_{c}\left(t\right)+ H_{\mathrm{gate}}\left(t\right)\right]\otimes\mathbb{I}_{E}+\mathbb{I}_{d}\otimes H_{E}+H_{\mathrm{int}},\label{eq:Htotal}
\end{equation}
where $H_{E}$ is the free Hamiltonian of the environment. In the picture obtained by unitarily
transforming Eq.~\eqref{eq:Htotal} using $U_{c}(t)$,
we obtain the Hamiltonian in what we henceforth call the control picture:
\begin{eqnarray}
&&H\left(t\right)  \equiv  \left[U_{c}^{\dagger}\left(t\right)\otimes\mathbb{I}_{E}\right]H_{\mathrm{tot}}\left(t\right)\left[U_{c}\left(t\right)\otimes\mathbb{I}_{E}\right]  \nonumber \\
&&+i\hbar\frac{dU_{c}^{\dagger}\left(t\right)}{dt}U_{c}\left(t\right) \otimes\mathbb{I}_{E} =  H_{G}\otimes\mathbb{I}_{E}+\mathbb{I}_{d}\otimes H_{E}\nonumber \\
 &  & +\left[U_{c}^{\dagger}\left(t\right)\otimes\mathbb{I}_{E}\right]H_{\mathrm{int}}\left[U_{c}\left(t\right)\otimes\mathbb{I}_{E}\right],\label{eq:H}
\end{eqnarray}
where, as we explain shortly, we have chosen $H_{\mathrm{gate}}(t)$
as
\begin{equation}
H_{\mathrm{gate}}\left(t\right) \equiv  U_{c}\left(t\right)H_{G}U_{c}^{\dagger}\left(t\right).\label{eq:H0}
\end{equation}
Note that at time $\tau$, the qudit state
in the control picture coincides with the one in the original picture
at the same time, explaining why we have chosen Eq.~\eqref{eq:H0}
and $t_0$ such that $\tau$ is an integer multiple of it. Since $H_{\mathrm{gate}}(t)$ is used in the presence of continuous dynamical decoupling, the evolution
proceeds as if only $U_{G}$ governed the qudit evolution
in the control picture (see Appendix \ref{Sec:Dynamical decoupling condition} for details). At time $\tau$, even in the original picture
the qudit state is then the one that the ideal noise-free evolution would produce,
up to a high-enough fidelity.

The dissipative dynamics is assumed to be resulting from a perturbing interaction
between the qudit and its environment described by the very general
Hamiltonian:
\begin{equation}
H_{\mathrm{int}}  =  \sum_{r=0}^{d-1}\sum_{s=0}^{d-1}\left|r\right\rangle \left\langle s\right|\otimes B_{r,s},\label{eq:Hint}
\end{equation}
where $B_{r,s}$, for $r,s=0,1,2,\ldots,d-1$ are operators that act
on the environmental states and $|k\rangle $, for $k=0,1,2,\ldots,d-1$,
are $d$ normalized state vectors forming a basis set for a Hilbert space of dimension $d$,
henceforth called the qudit space. This
is also to be considered the logical basis.
Notice that if the free Hamiltonian has not been eliminated before the application of the GCDD procedure, it can be thought as included in the above interaction Hamiltonian in such a way that the protection procedure in the end eliminates also the action of this free Hamiltonian.

\section{Our prescription}\label{Sec: Our prescription}

Now, we prescribe how to construct the required $U_{c}(t)$ and we address an arbitrary quantum gate operating on the qudit state,  showing how to protect its action against general noise. Also, we illustrate how to generate the fields in the laboratory whose action allows one to implement the control Hamiltonian $H_{c}(t)$ and the time-dependent gate Hamiltonian $H_{\mathrm{gate}}(t)$, the sum of which generates the evolution of the qudit state driven by such external fields.

\subsection{Constructing $U_{c}(t)$}\label{Sec:U_c}
We begin defining $H_{L}$
as the Hermitian operator whose action on the logical basis states
gives
\begin{equation}
H_{L}\left|k\right\rangle   \equiv  k\hbar\omega_{d}\left|k\right\rangle ,\label{eq:HL}
\end{equation}
for $k=0,1,\dots,d-1$, where
\begin{equation}
\omega_{d} \equiv  d \,\omega_{0}, \quad \omega_{0}  =  \frac{2\pi}{t_{0}},\label{omega0}
\end{equation}
being $\omega_{0}$ the control frequency corresponding to the dynamical-decoupling
period $t_{0}$.
The quantum-Fourier transform of the logical basis is given by \cite{BB01}
\begin{equation}
\left|\psi_{n}\right\rangle   \equiv  \frac{1}{\sqrt{d}}\sum_{j=0}^{d-1}\exp\left(\frac{2\pi i}{d}jn\right)\left|j\right\rangle ,\label{eq:psin}
\end{equation}
for $n=0,1,\dots,d-1$. We define the Hermitian operator $H_{F}$
by its action on the quantum-Fourier transformed basis, which is
\begin{equation}
H_{F}\left|\psi_{n}\right\rangle   \equiv  n\hbar\omega_{0}\left|\psi_{n}\right\rangle ,\label{eq:HF}
\end{equation}
for $n=0,1,\dots,d-1$. The required control unitary transformation
is then given by
\begin{equation}
U_{c}\left(t\right) \equiv \exp\left(-i\ome t\right)\exp\left(-i\frac{H_{L}}{\hbar}t\right)\exp\left(-i\frac{H_{F}}{\hbar}t\right),
\label{eq:Uc(t)}
\end{equation}
where we have defined a real constant $\ome$
as
\begin{equation}
\ome  \equiv - \frac{\mathrm{Tr}\left\{H_{L}\right\}+\mathrm{Tr}\left\{H_{F}\right\}}{\hbar d}.\label{eq:omegac}
\end{equation}

Using Eqs.~(\ref{eq:HL}-\ref{eq:Uc(t)}), it is now straightforward to obtain 
\begin{eqnarray}
&&\int_{0}^{t_{0}}dt\,U_{c}^{\dagger}\left(t\right)\left|p\right\rangle \left\langle q\right|U_{c}\left(t\right) \nonumber \\
&&  =  \frac{1}{d}\sum_{m=0}^{d-1}\sum_{n=0}^{d-1}\exp\left(-\frac{2\pi i}{d}pn\right)
\exp\left(\frac{2\pi i}{d}qm\right)  \left|\psi_{n}\right\rangle \left\langle \psi_{m}\right| \nonumber \\
&&\times
\int_{0}^{t_{0}}dt\, \exp\left[i\left(\left(n-m\right)+\left(p-q\right)d\right)\omega_{0}t\right] =  \frac{t_{0}}{d}\delta_{p,q}\mathbb{I}_{d},\nonumber \\ \label{eq:main}
\end{eqnarray}
where the time integration is not zero if and only if
$
(n-m)+(p-q)d =  0$. Given the constraints on the various parameters, this happens only if $m=n$ and $p=q$.
It
follows from Eqs.~\eqref{eq:Hint}  and \eqref{eq:main} that a weaker condition than Eq.~\eqref{eq:dd}
is obtained since the integral there appearing results to be equal to $\frac{t_{0}}{d}\mathbb{I}_{d} \otimes \sum_{p=0}^{d-1} B_{p,p}$. This weaker condition has the same form of Eq.~\eqref{conditionweak}. As said above, this condition is enough to obtain dynamical decoupling.

\subsection{The laboratory Hamiltonian}\label{Sec:The laboratory Hamiltonian}
To implement the GCDD, we need a prescription for $H_{c}(t)$
and $H_{\mathrm{gate}}(t)$. Using Eq.~\eqref{eq:Uc(t)}, we can calculate the control Hamiltonian,
$H_{c}(t)=i\hbar\frac{dU_{c}\left(t\right)}{dt}U_{c}^{\dagger}\left(t\right) $, as
\begin{equation}
H_{c}\left(t\right) =   \hbar\ome\mathbb{I}_{d}+H_{L}+U_{L}\left(t\right)H_{F}U_{L}^{\dagger}\left(t\right),\label{eq:Hc(t)}
\end{equation}
where, for simplicity, we have defined
$U_{L}\left(t\right)  \equiv  \exp(-iH_{L}t/\hbar)$.
Equation~\eqref{eq:H0} gives $H_{\mathrm{gate}}(t)$ in terms of
$U_{c}(t)$ and $H_{G}$. Hence, in the laboratory we need
to generate external fields such that they interact with the qudit
according to the following Hamiltonian:
\begin{eqnarray}
&&H_{\mathrm{lab}}\left(t\right)  \equiv H_{c}(t)+H_{\mathrm{gate}}(t) \nonumber \\
 && =     \hbar\ome\mathbb{I}_{d}+H_{L}+U_{L}\left(t\right)H_{F}U_{L}^{\dagger}\left(t\right) +U_{c}\left(t\right)H_{G}U_{c}^{\dagger}\left(t\right).\qquad\label{Hlab}
\end{eqnarray}
The term proportional to the unit matrix is immaterial to the dynamics,
since it only gives rise to a global phase factor multiplying the evolved state vector.

The action of the gate Hamiltonian $H_{G}$
is what we want to effectively perform, after the time interval
$\tau$, through $H_{\mathrm{lab}}\left(t\right)$. The Hamiltonian  $H_{G}$ can be expanded in the computational basis
by
\begin{equation}\label{Hd}
H_{G}  =  \hbar\sum_{r=0}^{d-1}\sum_{s=0}^{d-1}g_{r,s}\left|r\right\rangle \left\langle s\right| ,
\end{equation}
where $g_{s,r}^{\ast} =  g_{r,s}\label{grs}$ because $H_{G}$ is Hermitian. If some of the eigenvalues of $H_{G}$
are positive, let's choose, of these, the one with the highest absolute
value, say, $\hbar g_{0}$, with $g_{0}\geqslant0$, where the equal
sign is chosen if there are no positive eigenvalues of $H_{G}$. Then,
let's define the Hermitian operator $G$ such that
\begin{equation}
H_{G}  =  \hbar g_{0}\mathbb{I}_{d}-\hbar G.\label{HG}
\end{equation}
Therefore, $-G$ does not have positive eigenvalues and, thus, it's
a non-positive operator. Notice that $G$, however, is a non-negative
operator [we have introduced the minus sign appearing in Eq.~\eqref{HG}
just for convenience)].

Now, let us take a look at $H_{c}(t)$ of Eq.~\eqref{eq:Hc(t)}. 
It is easy to see, from Eqs.~\eqref{eq:HL} and \eqref{eq:HF}, that if
we define Hermitian operators $H_{L}^{\prime}$ and $H_{F}^{\prime}$
by
\begin{equation}
H_{L}^{\prime}  \equiv  \hbar\left(d-1\right)\omega_{d}\mathbb{I}_{d}-H_{L}\label{HLprime},
\end{equation}
and
\begin{equation}
H_{F}^{\prime}  \equiv  \hbar\left(d-1\right)\omega_{0}\mathbb{I}_{d}-H_{F},\label{HFprime}
\end{equation}
respectively, then $H_{L}^{\prime}$ and $H_{F}^{\prime}$ are both
non-negative operators. Now, using Eqs.~{\eqref{HG}, \eqref{HLprime}, and \eqref{HFprime},
we obtain from Eq.~\eqref{Hlab} that
\begin{eqnarray}
&&H_{\mathrm{lab}}\left(t\right) =  \hbar\left[g_{0}+\omega_{r}+\left(d^{2}-1\right)\omega_{0}\right]\mathbb{I}_{d}  -\bigg[H_{L}^{\prime}   \nonumber \\
 &  &  +\exp\left(-i\frac{H_{L}}{\hbar}t\right)H_{F}^{\prime}\exp\left(i\frac{H_{L}}{\hbar}t\right)+\hbar U_{c}\left(t\right)GU_{c}^{\dagger}\left(t\right)\bigg],\nonumber \\ \label{Hlab(t)}
\end{eqnarray}
where we have also used Eq.~\eqref{omega0}.
Since a unitary transformation
of a non-negative operator is still a non-negative operator and $H_{L}^{\prime}$,
$H_{F}^{\prime}$, and $G$, as we have defined them, are all non-negative,
it follows that the last term within square brackets on the right-hand
side of Eq.~\eqref{Hlab(t)} is a non-negative operator. Then, we
can define
\begin{equation}
\Upsilon\left(t\right)  \equiv  \sqrt{\frac{V}{\hbar}}\label{Upsilon} ,
\end{equation}
where
 \begin{eqnarray}
 V&=&  H_{L}^{\prime}+  \exp\left(-i\frac{H_{L}}{\hbar}t\right)  H_{F}^{\prime}\exp\left(i\frac{H_{L}}{\hbar}t\right)   \nonumber  \\
 & & +\,
\hbar U_{c}\left(t\right)GU_{c}^{\dagger}\left(t\right),
\end{eqnarray}
and rewrite Eq.~\eqref{Hlab(t)} as
\begin{equation}
H_{\mathrm{lab}}\left(t\right)  =  \hbar\omega_{l}\mathbb{I}_{d}-\hbar\Upsilon\left(t\right)\Upsilon\left(t\right),\label{Hlab-Upsilon}
\end{equation}
where
\begin{equation}
\omega_{l}  \equiv  g_{0}+\omega_{r}+\left(d^{2}-1\right)\omega_{0}.\label{omegag}
\end{equation}
Equation~\eqref{Hlab-Upsilon} gives the explicit laboratory Hamiltonian that can be used to implement the protective scheme. Also, it is important to emphasize that once the target state of the gate is achieved, thanks to our GCDD procedure, we can preserve the final state in the absence of the gate but under the noise. This is done by simply turning the protection on, without the control fields that generated the gate operation (this is obtained by using  $H_G=0$). In this way the memory of the final state is protected.

In the next section, we apply the GCDD method to the case of a
qutrit based on the three magnetic hyperfine states of the ground
energy level of $^{87}\mathrm{Rb}$. However, in principle, we could use the same kind of two-photon atomic
transitions, employed to implement this qutrit, and involve any number of Zeeman hyperfine states. Thus,
although the model implementation we use is limited to the simple
case of the ground state hyperfine states of the $^{87}\mathrm{Rb}$
atom, other atomic systems could be used to obtain control over systems
of qudits of dimension $d=2$ or $d>3$. Let us, then, proceed with the illustration
of the application of the GCDD method.

\section{Application of the GCDD}\label{Sec: atomic qutrit}
To illustrate the application of the GCDD method, we describe a possible implementation
of a  qutrit, exploiting the three magnetic hyperfine states of the ground
energy level of $^{87}\mathrm{Rb}$. In the following, we use atomic data from Ref.~\cite{NIST}. Figure~\ref{fig:D2line} shows
the relevant $\mathrm{D_2}$-line hyperfine states of $^{87}\mathrm{Rb}$.
In the following, we show that this model allows one to generate continuously all the possible SU(3) group operations which are, in general, needed to apply the GCDD procedure. 

\begin{figure}[t!]
\centering
\includegraphics[width=0.46\textwidth]{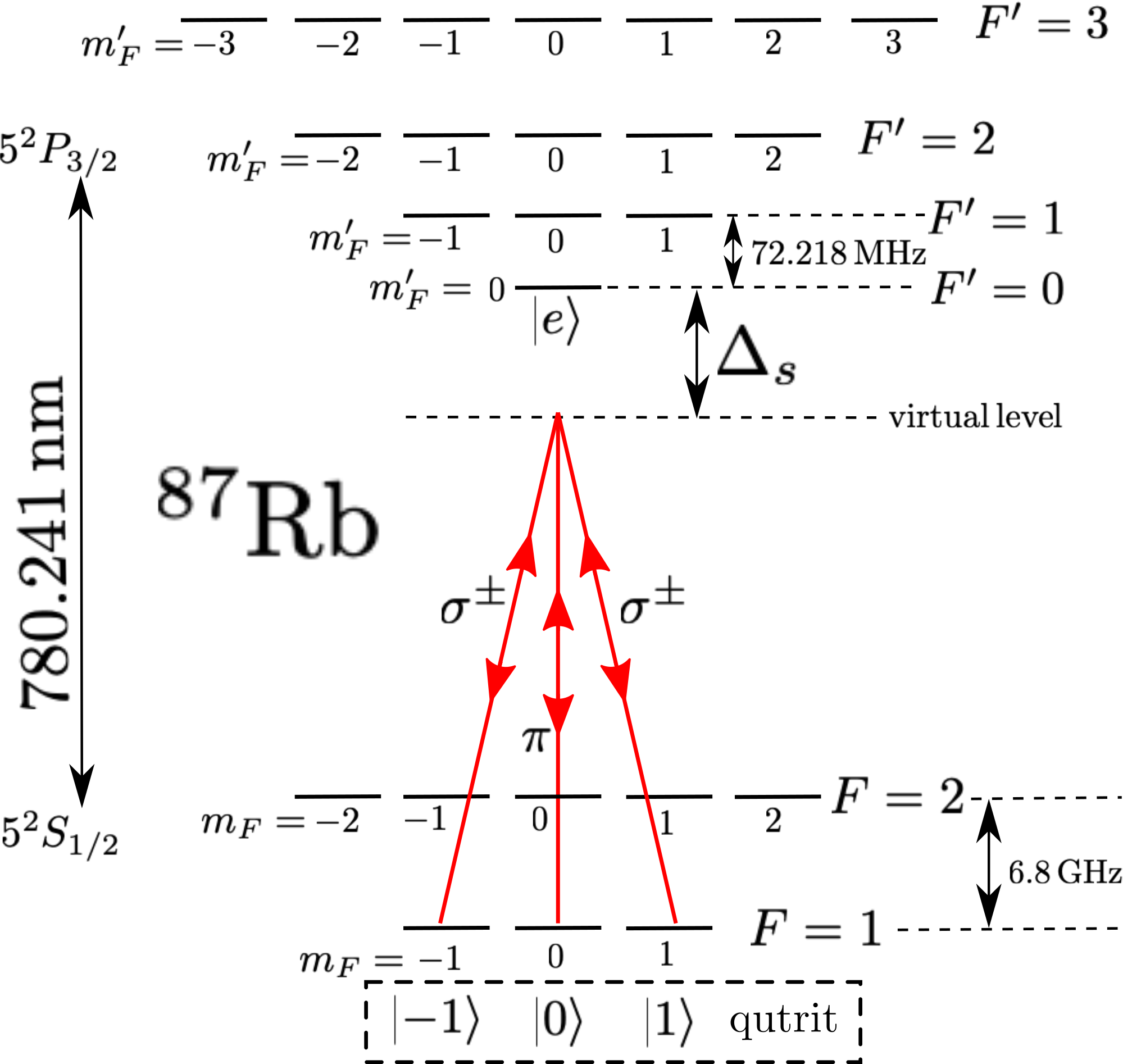}
\caption{\label{fig:D2line}The hyperfine degenerate states of the $\mathrm{D} _{2}$ transition of $^{87}\mathrm{Rb}$ (not to scale). We are using two-photon transitions for three different detunings $\Delta _{s}$, for $s=1,2,3$, and, for each of these detunings, we use $\sigma ^{\pm}$- and $\pi $-polarized laser light. The wavelength 780.241 nm corresponds to a frequency of the order of 384.230 THz. The qutrit space comprises the subspace spanned	by the three magnetic states of the $F=1$ ground level of $^{87}\mathrm{Rb}$,
with magnetic quantum numbers $m_{F}=-1,0,1$. We represent these degenerate
states by the kets $|m \rangle$, for $m=-1,0,1$, respectively. }
\end{figure}
The $^{87}\mathrm{Rb}$ atom, in the absence of external magnetic
fields, has a ground-state manifold of three degenerate magnetic states.
This is so because $^{87}\mathrm{Rb}$ has a nuclear spin equal to
$3/2$ and a fundamental electronic manifold of states with symmetry
$5^{2}S_{1/2}$. This amounts to a hyperfine ground state with total angular momentum
$F=1$, so that there are three magnetic states whose projections
along the quantization axis have quantum numbers $m_{F}=-1,0,1$.
These three ground states are degenerate in the absence of magnetic
fields and we denote them by $|m\rangle$, for $m=-1,0,1$ (the notations $1$ and $+1$ are both used in the following).
The $5^{2}S_{1/2}$ ground manifold of states (including also the
five magnetic states of the $F=2$ ground level, besides the already-mentioned
three $F=1$ states) can be excited to states of the $5^{2}P_{3/2}$
excited manifold by absorbing photons with wavelengths of about $780$
nm (called the $\mathrm{D}_{2}$ spectral line of $^{87}\mathrm{Rb}$).
The lowest-energy magnetic hyperfine state of the $5^{2}P_{3/2}$
manifold is not degenerate and has a total-angular-momentum quantum
number $F^{\prime}=0$, whose projection is $m_{F}^{\prime}=0$. We denote this state by $|e\rangle$  and we name $\hbar\omega_{g}$ the energy of the ground states $|m\rangle$, for $m=-1,0,1$, and  $\hbar\omega_{e}$ the energy of the excited state $|e\rangle$. 

If we use only  frequencies
corresponding to virtual transitions with wavelengths greater than
the optical $780$ nm, that is, if we use only photons with frequencies $\omega_s$ that are red-detuned
from the $F=1\leftrightarrow F^{\prime}=0$
transition (this means that the detunings $\Delta_{s}  \equiv  \omega_{s}-\omega_{e}+\omega_{g}$ are negative), then we can approximate the relevant set of atomic states
to be the one involving only the ground states $|m\rangle $,
for $m=-1,0,1$, and the excited state $|e\rangle $. For the restricted Hilbert space of these four atomic hyperfine states, denoted by $\mathscr{H}_{4}$,
we have the identity operator
\begin{equation}
\mathbb{I}_{4} =  \sum_{m=-1}^{1}\left|m\right\rangle \left\langle m\right|+\left|e\right\rangle \left\langle e\right|.\label{I4}
\end{equation}
If the photons are detuned far enough to the red of the transitions
$|m\rangle \leftrightarrow |e\rangle$, for
$m=-1,0,1$, then the excited state is not going to be effectively
populated, avoiding spurious transitions to the $F=2$ ground states
through spontaneous emission from $|e\rangle$. The effective
qutrit, therefore, consists of the states $|m\rangle$,
with $m=-1,0,1$, whose Hilbert space we denote by $\mathscr{H}_{3}$.

As explained in Appendix~\ref{App:GCDD for an atomic qutrit}, the control over the states in $\mathscr{H}_{3}$ using the GCDD method
is accomplished through two-photon transitions. These transitions are used to couple these states among themselves in a controlled way. In particular, in order to be able to generate continuously all the possible SU(3) group operations we need three independent detunings $\Delta_{s}$, with $s=1,2,3$, one for each  transition
$|m\rangle \leftrightarrow |e\rangle$, for
$m=-1,0,1$. In particular, each laser beam is red-detuned from $|e\rangle$, i.e., $\Delta_{s}<0$ for $s=1,2,3$.
For each of the three different laser colors we can use
the linear polarization and both circular polarizations, thus obtaining
a total of nine independent Rabi frequencies. These independent control
parameters are enough to effectively emulate the action of any $3\times3$ Hermitian
matrix used to represent a generic modified single-qutrit quantum gate, $H_{\mathrm{gate}}(t)$ [cf. Eq.~\eqref{eq:H0}],
together with the control fields described by $H_{c}(t)$,
that are required for the GCDD, as explained in detail in Appendix~\ref{App:GCDD for an atomic qutrit}. There, we present all the
details and values of the parameters that we can use, in principle, to implement  the above atomic qutrit and the needed effective control Hamiltonian. Here, it suffices to say
that using the rotating-wave
approximation and adiabatic elimination of $|e\rangle $
\cite{Raman}, we get an effective Hamiltonian based on two-photon interactions given by [cf. Eq.~\eqref{OmegadaggerOmega}]
\begin{multline}
H_{\mathrm{eff}}\left(t\right) = \hbar \omega_g \mathbb{I}_{3}  \\- \hbar\sum_{m=-1}^{1}\sum_{n=-1}^{1}\sum_{s=1}^{3}\frac{\Omega_{s,-n}^{\ast}\left(t\right)\Omega_{s,-m}\left(t\right)}{\Delta_{s}}\left|m\right\rangle \left\langle n\right|,
	\label{Heff}
\end{multline}
where $\Omega_{s,-m}(t)$, for $s=1,2,3$ and $m=-1,0,1$,
are adiabatically time-varying Rabi frequencies allowing one to emulate the time dependent
control Hamiltonian of Eq.~\eqref{Hlab} up to an immaterial term proportional to $\mathbb{I}_{3}$ [cf. Eq.~\eqref{Hlab-Upsilon}] (this term, which is immaterial for the dynamics, is not even needed if the atomic energies are shifted in such a way that $\omega_g=\omega_l$).

As quantum gate to implement for the above qutrit we choose the Hadamard one. Starting from the ideal definition of the Hadamard unitary quantum gate \cite{qutritHadamard} in the ground-state subspace basis \{$|-1\rangle$, $|0\rangle$, $|1\rangle$\},
namely,
\begin{equation}
U_{G}  =  \frac{1}{i\sqrt{3}}\left[\begin{array}{ccc}
1 & 1 & 1\\
1 & \exp\left(2\pi i/3\right) & \exp\left(4\pi i/3\right)\\
1 & \exp\left(4\pi i/3\right) & \exp\left(2\pi i/3\right)
\end{array}\right],\label{Hadamard-unitary}
\end{equation}
we can invert the equation
\begin{equation}
U_{G}  =  \exp\left(-i\frac{H_{G}}{\hbar}\tau\right),\label{UG}
\end{equation}
to obtain the gate Hamiltonian:
\begin{eqnarray}
H_{G} & = & \frac{\pi\hbar}{4\sqrt{3} \tau}\left[\begin{array}{ccc}
4\sqrt{3}-2 & -2 & -2\\
-2 & 2\sqrt{3}+1 & 2\sqrt{3}+1\\
-2 & 2\sqrt{3}+1 & 2\sqrt{3}+1
\end{array}\right],\label{HHadamard}
\end{eqnarray}
where $\tau$ is the characteristic gate time. For example, starting from the state $| 0 \rangle $, after the action of the Hadamard operation, the output state after a time $\tau$ becomes
\begin{equation}
 \left|\psi \right\rangle  =  \displaystyle -\frac{i}{\sqrt{3}}\left[\left|-1\right\rangle + \exp\left(i\varphi \right)\left|0\right\rangle +\exp\left(- i\varphi \right)\left|1\right\rangle \right],\label{FinalPsi}
\end{equation}
where $\varphi=2\pi/3$. 

In the next section, in our numerical simulations, we consider two paradigmatic noises due to bosonic thermal baths, chosen to disturb our intended gate operation. We consider the amplitude damping noise which simulates dissipation involving, respectively, $| - 1\rangle$ and $| 0\rangle $, and   $| 1\rangle$ and $| 0\rangle $, and the dephasing noise which destroys the relative coherences among these states.

\section{Numerical simulations}\label{numsim}

In this section we present our numerical simulations in the case of the qutrit depicted in Fig.~\ref{fig:D2line} when both damping and dephasing are simultaneously present. In particular, we perform our simulations by following the prescription of Sec.~\ref{Sec: Our prescription} and we assume the presence of two identical baths with Ohmic spectral density, characterized by an exponential cutoff function with an angular cutoff frequency $\omega_c=2\pi/\tau_c$, where  $\tau_c$ is the  bath correlation time.
The effectiveness of our protective scheme is studied by means of the fidelity measuring how close are the states obtained, respectively, by the dissipative dynamics induced by Eq.~\eqref{eq:Htotal} (treated by means of a Redfield master equation as explained below) and by the noise-free dynamics governed by the Hamiltonian $H_{\mathrm{lab}}(t)$ of Eq.~\eqref{Hlab}. We recall that this latter dynamics
gives after a time $\tau$ the same output state of the ideal noise-free dynamics governed by the Hadamard gate Hamiltonian $H_G$ of Eq.~\eqref{HHadamard} (i.e., the case without protective scheme). The dissipative dynamics in the absence of the protective scheme is obtained by turning off the control Hamiltonian $H_c(t)$. The fidelity is defined for two arbitrary states $\rho$ and $\sigma$ as $[\mathrm{Tr} \{\sqrt{\sqrt{\rho} \sigma \sqrt{\rho}}t\}]^2 $. Here, the open quantum system dynamics is obtained by means of the Redfield master equation which describes the dissipative dynamics in the general case. In the Appendix \ref{Sec:noise} we show in details how to calculate the dynamics governed by this master equation.

\begin{figure}[t!] \centering \includegraphics[width=0.46\textwidth] {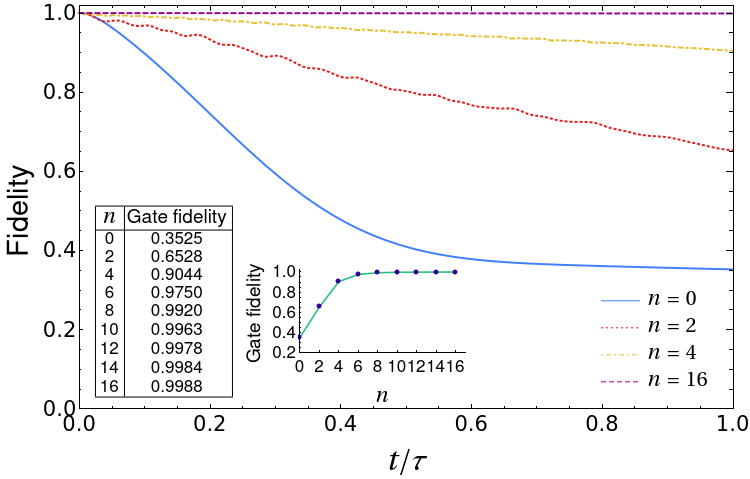} \caption{\label{fig:protection} Numerical solutions for the GCDD to overcome amplitude damping and dephasing during the action of the Hadamard gate (the time $t$ is in units of the gate time $\tau$), with $| 0\rangle $ as initial state. Here,  the bath correlation time is $\tau_c=\tau/4$,   $\hbar \omega_c/(k_B T)=1$, and the involved coupling constants are all equal leading to an effective coupling parameter $\bar{\lambda}=0.1$  (see Appendix \ref{Sec:noise} for details).  The solid line represents the fidelity with no protection, while the dotted, dot-dashed, and dashed ones refer to the protective scheme  with $\omega_0/(2\pi)$ equal to $2/\tau_c$, $4/\tau_c$, and $16/\tau_c$, respectively. In the insets, we represent the gate fidelity (fidelity at time $\tau$) as a function of $n=\omega_0\tau_c/(2\pi)$ (the interpolated curve just guides the reading) and we report its numerical values. } \end{figure}

Figure~\ref{fig:protection} shows that if we do not use the GCDD method during the time $\tau$ in which the Hadamard gate operates and let the noise affect the dynamics, starting
from the state $|0\rangle$, the fidelity rapidly decreases. 
When the GCDD protection is turned on, we obtain better results by increasing the control frequency $\omega_0$, with the final gate fidelity moving towards one. In the insets, we show the gate fidelity, i.e., the fidelity at time $\tau$, as a
function of $n=\omega_0 \tau_c/(2\pi)$ [this implies $t_{0}=\tau_c/n$, according to Eq.~\eqref{omega0}] and we report its numerical values. 
 In particular, the smaller  is $t_0$ with respect to $\tau_c$, the more effective is the decoupling procedure. We stress out that, by construction, the time at which to look for a state close to the original target is exactly the time $\tau$ at which the original gate would have produced that state in the absence of the environment and of the control fields. The actual value of the gate time $\tau$ is not specified in these simulations, the other quantities being given in units of it. Its value must just be such that the derivation of the effective Hamiltonian of Eq.~\eqref{Heff} can be coherently performed. We state this condition for a question of coherence in our analysis, even if in our simulations we do not make use of Eq.~\eqref{Heff}, but we directly follow the prescriptions given in Sec.~\ref{Sec: Our prescription}. We also observe that the results shown in Fig.~\ref{fig:protection} have been obtained when the various coupling constants involved in the interaction Hamiltonian with the environment are all equal leading to an effective coupling parameter $\bar{\lambda} = 0.1$ (see Appendix \ref{Sec:noise} for details). We have also tested some configurations with the various coupling constants not all equal, finding similar results.

The illustration of the GCDD method
shown in Fig.~\ref{fig:protection} for our qutrit model using $^{87}\mathrm{Rb}$ and
a modulated set of laser beams, can, in principle, be realistically
implemented in the laboratory.  Even if only
the qutrit case has been considered, our results show that quantum computation
could be implemented using laser light and atomic systems, which
are available in setups with trapped ions, for example. This kind of implementation
is attractive because it already presents long coherence times, implying
high efficiency of our procedure \cite{Raman,Treulein,Deutsch}. We remark that the implementation of our procedure can, in principle,  be extended to the
case of a qudit with more levels.

\section{Extension to the case of a multi-qudit scenario}\label{Sec: Extension to the case of a multi-qudit scenario}

The domain of applicability of our GCDD procedure can be {extended} to the case one wants to protect the action of a multi-qudit gate on an ensemble of qudits, identical or not, subject to general noise, which can act locally or non-locally on them. This can be obtained by extending the procedure presented in detail in Appendix~\ref{SeveralQudits} for the case of two qudits. There, it is explicitly shown how to build up the control fields necessary to protect the qudits from the action of a general noise. The extension to the case of more than two qudits, identical or not, is straightforward, as indicated in  Appendix~\ref{SeveralQudits}.

A simple but important  direct application of this procedure is the possibility to preserve the memory of a multi-qudit entangled state against noise.

\section{Conclusions}
In conclusion, here we have presented a generalized continuous dynamical decoupling procedure to decouple an  ensemble of 
qudits from any possible noise and
still apply an arbitrary many-qudit quantum gate on them.
Our study  extends the domain of applicability of dynamical decoupling strategies. Indeed, we provide a general procedure of this kind to protect the action of a general quantum gate on one or many qudits, identical or not,  against general noise.

Importantly, we have explicitly provided a detailed analysis of a specific example, employing a Rubidium atom, which, in principle, could be experimentally implemented, where it is explicitly shown how to realize the operations which could be in general needed to apply our GCDD procedure.

We think then that our approach represents an important step towards the protection of quantum information, especially when many-level quantum systems are employed. We stress that our procedure appears to be directly implementable in experiments with atomic systems, nitrogen-vacancy centers, or other setups where current technology permits to generate the control fields required for the protection scheme.

\section*{Acknowledgments}

R.d.J.N. acknowledges support from Funda\c{c}\~{a}o de Amparo \`{a} Pesquisa
do Estado de S\~{a}o Paulo (FAPESP), project number 2018/00796-3, and
also from the National Institute of Science and Technology for Quantum
Information (CNPq INCT-IQ 465469/2014-0) and the National Council for Scientific and
Technological Development (CNPq). F.F.F acknowledges support from Funda\c{c}\~{a}o de Amparo \`{a} Pesquisa
do Estado de S\~{a}o Paulo (FAPESP), project number 2019/05445-7, and thanks the Institut UTINAM  for its hospitality and financial support during his visit in Besan\c{c}on. B.B. acknowledges  support by the French ``Investissements d'Avenir'' program, Project ISITE-BFC (Contract No. ANR-15-IDEX-03).

\appendix

\section{Dynamical decoupling condition}\label{Sec:Dynamical decoupling condition}
	
Here, we show that dynamical decoupling is still obtained if a weaker condition than the one of Eq.~\eqref{eq:dd}  is satisfied.
 
Let us start observing that Eq.~\eqref{eq:dd} is stronger than necessary to achieve dynamical decoupling. A weaker, but sufficient, condition would be that, instead of being
equal to zero, the integral in Eq.~\eqref{eq:dd} results in a tensor product
between the qudit identity and an environment operator. To see this,
we recall that the idea of dynamical decoupling comes from the usual
Magnus expansion \cite{Facchi} of the total propagator for the whole system, the qudit and
the environment, that is,
\begin{equation}
U\left(\tau\right) \approx \exp\left(-\frac{i}{\hbar}\left\langle H\right\rangle \tau\right),\label{Utot-result}
\end{equation}
where we have defined the time average of $H(t)$ as
\begin{equation}
\left\langle H\right\rangle   \equiv  \frac{1}{t_{0}}\int_{0}^{t_{0}}ds_{1}H\left(s_{1}\right)\label{averageH},
\end{equation}
and $H(t)$ is the total Hamiltonian given in Eq.~\eqref{eq:H}. Hence, we obtain:
\begin{eqnarray}
\left\langle H\right\rangle  & = & H_{G}\otimes\mathbb{I}_{E}+\mathbb{I}_{d}\otimes H_{E} \nonumber\\&&
+\,\frac{1}{t_{0}}\int_{0}^{t_{0}}dt\left[U_{c}^{\dagger}\left(t\right)\otimes\mathbb{I}_{E}\right]H_{\mathrm{int}}\left[U_{c}\left(t\right)\otimes\mathbb{I}_{E}\right].\qquad\label{<H>}
\end{eqnarray}
Equation~\eqref{Utot-result} is not exact because, as usual \cite{Facchi}, we have neglected terms of order equal and superior to $t_{0}$ in the Hamiltonian part appearing in the argument of the exponential. However, it becomes exact if the number of periods within
the time interval $\tau$ tends to infinity. We now assume that a weaker condition than the one of Eq.~\eqref{eq:dd} is satisfied, namely [the same condition given in Eq.~\eqref{conditionweak}],
\begin{equation}
\int_{0}^{t_{0}}dt\left[U_{c}^{\dagger}\left(t\right)\otimes\mathbb{I}_{E}\right]H_{\mathrm{int}}\left[U_{c}\left(t\right)\otimes\mathbb{I}_{E}\right] = c \, t_{0} \,\mathbb{I}_{d}\otimes B,\label{condition}
\end{equation}
where $c$ is a constant and $B$ is some operator that acts on the
environment. It follows that Eq.~\eqref{<H>} becomes:
\begin{equation}
	\left\langle H\right\rangle  = H_{G}\otimes\mathbb{I}_{E}+\mathbb{I}_{d}\otimes \left( H_{E}+c\, B \right).
\end{equation}
Therefore, with this average Hamiltonian, Eq.~\eqref{Utot-result}
gives
\begin{eqnarray}
	&&U\left(\tau\right)  \approx  \exp\left[-\frac{i}{\hbar}\left(H_{G}\otimes\mathbb{I}_{E}+\mathbb{I}_{d}\otimes \left(H_{E}+c\, B\right)\right)\tau\right] \nonumber\\
	&  &= \exp\left[-\frac{i}{\hbar}\left(\mathbb{I}_{d}\otimes \left( H_{E}+c\, B\right)\right)\tau\right]\exp\left[-\frac{i}{\hbar}\left(H_{G}\otimes\mathbb{I}_{E}\right)\tau\right],\nonumber\\
\end{eqnarray}
which shows that the qudit is decoupled from the environment, since the interactions get effectively eliminated, even in the case in
which $c\neq0$ in Eq.~\eqref{condition}.

\section{GCDD for an atomic qutrit}
\label{App:GCDD for an atomic qutrit}

In this Appendix we show how to effectively implement an atomic qutrit with the system described in Sec.~\ref{Sec: atomic qutrit} and how to realize within this system an effective control Hamiltonian like  the one of Eq.~\eqref{Hlab} (in general up to an immaterial term proportional to $\mathbb{I}_{3}$), needed to apply our GCDD procedure.

\subsection{The interaction Hamiltonian between the atom and the laser beams}

We introduce nine laser beams, whose electric-field vectors, each
being the resultant with a different polarization, can be written as
\begin{eqnarray}
\mathbf{E}_{\pm1}\left(t\right) & = & \sum_{s=1}^{3}\left[\mathscr{E}_{s,\pm1}\left(t\right)\boldsymbol{\hat{\varepsilon}}_{\pm1}\exp\left(-i\omega_{s}t\right) \right. \nonumber \\
& & \left. + \mathscr{E}_{s,\pm1}^{\ast}\left(t\right)\boldsymbol{\hat{\varepsilon}}_{\pm1}^{\ast}\exp\left(i\omega_{s}t\right)\right]
\label{Epm}
\end{eqnarray}
and
\begin{equation}
\mathbf{E}_{0}\left(t\right)  =  \boldsymbol{\hat{\varepsilon}}_{0}\sum_{s=1}^{3}\left[\mathscr{E}_{s,0}\left(t\right)\exp\left(-i\omega_{s}t\right)+\mathscr{E}_{s,0}^{\ast}\left(t\right)\exp\left(i\omega_{s}t\right)\right], 
\label{E0}
\end{equation}
where the polarization versors are chosen, in terms of a space-fixed
system of Cartesian coordinates, as
\begin{equation}
\boldsymbol{\hat{\varepsilon}}_{\pm1}  \equiv  \mp\left(\frac{\mathbf{\hat{x}}\pm i\mathbf{\hat{y}}}{\sqrt{2}}\right),\label{epsilonpm}
\end{equation}
representing, respectively, the $\sigma^{\pm}$ polarizations, and
\begin{equation}
\boldsymbol{\hat{\varepsilon}}_{0}  \equiv  \mathbf{\hat{z}},\label{epsilon0}
\end{equation}
representing the $\pi$ polarization. Here, the $z$-axis of this
system is chosen to represent the quantization axis. It is noteworthy
that in Eqs.~\eqref{Epm} and \eqref{E0}, for each polarization,
there are three different superposed amplitudes, $\mathscr{E}_{s,\pm}(t)$
and $\mathscr{E}_{s,0}(t)$,  each corresponding to a different polarization-independent frequency, $\omega_{s}$, for $s=1,2,3$. Figure~1 shows the scheme we are describing. The amplitudes $\mathscr{E}_{s,\pm}(t)$ and $\mathscr{E}_{s,0}(t)$, as we discuss below, must follow a prescribed relatively slow time-dependent	modulation. It is worth mentioning that we treat the driving electric fields of Eqs.~\eqref{Epm} and \eqref{E0} as classical, intense
laser fields. We are justified to use such a semiclassical approach because of the relatively high intensities and detuning magnitudes used, so that quantum fluctuations of the number of photons is completely negligible in the regime we consider here.

The laser beams of Eqs.~\eqref{Epm} and \eqref{E0} interact with
the atom according to the Hamiltonian
\begin{eqnarray}
H_{\mathrm{int}}\left(t\right) & = & -\mathbf{d}\boldsymbol{\cdot}\left[\mathbf{E}_{-1}\left(t\right) +\mathbf{E}_{0}\left(t\right)+ \mathbf{E}_{+1}\left(t\right)\right]\nonumber \\
& = & -\mathbf{d}\boldsymbol{\cdot}\sum_{q=-1}^{+1}\mathbf{E}_{q}\left(t\right),\label{dE}
\end{eqnarray}
since we have the three resultant laser fields continuous and simultaneously
present, each one with a different polarization, where $\mathbf{d}$
is the atomic electric-dipole operator, which is Hermitian. In Cartesian
coordinates, we write
\begin{equation}
\mathbf{d} = d_{x}\mathbf{\hat{x}}+d_{y}\mathbf{\hat{y}}+d_{z}\mathbf{\hat{z}}
\end{equation}
and, using Eq.~\eqref{I4} in $\mathbf{d}  =  \mathbb{I}_{4}\mathbf{d}\mathbb{I}_{4}$, we obtain
\begin{eqnarray}
\mathbf{d} &  = & \sum_{m=-1}^{1}\left|e\right\rangle \left\langle e\right|\mathbf{d}\left|m\right\rangle \left\langle m\right|+\sum_{m=-1}^{1}\left|m\right\rangle \left\langle m\right|\mathbf{d}\left|e\right\rangle \left\langle e\right|\nonumber \\
& = & \sum_{m=-1}^{1}\left|e\right\rangle \left(\left\langle m\right|\mathbf{d}\left|e\right\rangle \right)^{\ast}\left\langle m\right|+\sum_{m=-1}^{1}\left|m\right\rangle \left\langle m\right|\mathbf{d}\left|e\right\rangle \left\langle e\right|, \nonumber \\\label{I4dI4}
\end{eqnarray}
where we have used the fact that the electronic excited state has
a parity that is opposite to the parity of the ground states, that is,
\begin{equation}
\left\langle m\right|\mathbf{d}\left|m^{\prime}\right\rangle   =  \mathbf{0} \quad
\mathrm{and} \quad \left\langle e\right|\mathbf{d}\left|e\right\rangle   =  \mathbf{0},
\end{equation}
for $m, m^{\prime}=-1, 0, 1$. Now, we can write the operator $\mathbf{d}$
in terms of its spherical-tensor components:
\begin{eqnarray}
\mathbf{d} & = & d_{x}\mathbf{\hat{x}}+d_{y}\mathbf{\hat{y}}+d_{z}\mathbf{\hat{z}} \nonumber \\
& = & \frac{d_{x}-id_{y}}{\sqrt{2}}\boldsymbol{\hat{\varepsilon}}_{-1}^{\ast}+d_{0}\boldsymbol{\hat{\varepsilon}}_{0}-\frac{d_{x}+id_{y}}{\sqrt{2}}\boldsymbol{\hat{\varepsilon}}_{+1}^{\ast} \nonumber\\
& = & d_{-1}\boldsymbol{\hat{\varepsilon}}_{-1}^{\ast}+d_{0}\boldsymbol{\hat{\varepsilon}}_{0}
+d_{+1}\boldsymbol{\hat{\varepsilon}}_{+1}^{\ast},
\end{eqnarray}
that is,
\begin{equation}
\mathbf{d} =  \sum_{q=-1}^{+1}d_{q}\boldsymbol{\hat{\varepsilon}}_{q}^{\ast},\label{dipole}
\end{equation}
where we have used Eqs.~\eqref{epsilonpm} and \eqref{epsilon0} and
defined its spherical components as usual:
\begin{equation}
d_{\pm1}  \equiv  \mp\frac{d_{x}\pm id_{y}}{\sqrt{2}}\label{dpm}
\end{equation}
and
\begin{equation}
d_{0}  \equiv  d_{z}.\label{d0}
\end{equation}

Because $|e\rangle $ has zero angular momentum, from Eq.~\eqref{dipole} it follows that
\begin{equation}
\left\langle m\right|\mathbf{d}\left|e\right\rangle   =  \sum_{q=-1}^{+1}\left\langle m\right|d_{q}\left|e\right\rangle \boldsymbol{\hat{\varepsilon}}_{q}^{\ast} =  \left\langle m\right|d_{m}\left|e\right\rangle \boldsymbol{\hat{\varepsilon}}_{m}^{\ast},\label{mde}
\end{equation}
since total angular momentum is conserved. From the Wigner-Eckart theorem \cite{wigner-eckart}, we have
\begin{equation}
\left\langle m\right|d_{q}\left|e\right\rangle  =  \delta_{q,m}D,\label{WET}
\end{equation}
where $D$ is a reduced matrix element of the dipole operator and
is, thus, independent of $m$ or $q$. Hence, we rewrite Eq.~\eqref{mde},
using Eq.~\eqref{WET}, as
\begin{equation}
\left\langle m\right|\mathbf{d}\left|e\right\rangle   =  D\boldsymbol{\hat{\varepsilon}}_{m}^{\ast}.\label{mde-1}
\end{equation}
Substituting Eq.~\eqref{mde-1} into Eq.~\eqref{I4dI4}, we obtain
\begin{equation}
\mathbf{d}  = \sum_{m=-1}^{+1}D^{\ast}\boldsymbol{\hat{\varepsilon}}_{m}\left|e\right\rangle \left\langle m\right|+\sum_{m=-1}^{+1}D\boldsymbol{\hat{\varepsilon}}_{m}^{\ast}\left|m\right\rangle \left\langle e\right|.\label{d}
\end{equation}
Substituting Eqs.~\eqref{Epm}, \eqref{E0}, and \eqref{d} into Eq.~\eqref{dE} gives
\begin{eqnarray}
H_{\mathrm{int}}&&\left(t  \right)  =  -\!\!\sum_{q=-1}^{+1}\sum_{s=1}^{3}\left(-1\right)^{q}D^{\ast}\mathscr{E}_{s,q}\left(t\right)\exp\left(-i\omega_{s}t\right)\left|e\right\rangle \left\langle -q\right|\nonumber \\
&  & -\sum_{q=-1}^{+1}\sum_{s=1}^{3}\left(-1\right)^{q}D\mathscr{E}_{s,q}^{\ast}\left(t\right)\exp\left(i\omega_{s}t\right)\left|-q\right\rangle \left\langle e\right|,\label{Hint(t)}
\end{eqnarray}
where we have used the rotating-wave approximation \cite{RWA}, which is justified since we will choose detunings $\Delta_{s} = \omega_{s}-\omega_{e}+\omega_{g}$, for $s=1,2,3$, and Rabi frequencies 
$\Omega_{s,q}(t)$, defined as
\begin{equation}
\Omega_{s,q}\left(t\right) \equiv \frac{1}{\hbar} \left(-1\right)^{q}D^{\ast}\mathscr{E}_{s,q}\left(t\right),\label{Rabi}
\end{equation}
for $s=1,2,3$ and $q=-1,0,+1$,	
much smaller in modulus than the transition frequency $\omega_{e}-\omega_{g}$ and the laser frequencies $\omega_s$. We have also used Eqs.~\eqref{epsilonpm} and \eqref{epsilon0}
to calculate the scalar products between polarization vectors. 
Values of $|\Omega_{s,q}(t)|/2\pi$
of the order of a few MHz, let us say, roughly
\begin{equation}
\frac{|\Omega_{s,q}\left(t\right)|}{2\pi}  \sim 1\textrm{ MHz},\label{orderOmega}
\end{equation}
are routinely obtained in the context of optical manipulation
of rubidium \cite{Treulein,Deutsch}. These independent control parameters
are enough to emulate the action of any $3\times3$ Hermitian matrix
used to represent a generic modified single-qutrit quantum gate [cf. Eq.~\eqref{eq:H0}] together with
the control fields required for the continuous dynamical
decoupling, as we explain below. Given Eq.~\eqref{Rabi}, we can rewrite
Eq.~\eqref{Hint(t)} as
\begin{eqnarray}
H_{\mathrm{int}}\left(t\right) & = & -\hbar\sum_{q=-1}^{+1}\sum_{s=1}^{3}\Omega_{s,q}\left(t\right)\exp\left(-i\omega_{s}t\right)\left|e\right\rangle \left\langle -q\right|\qquad\nonumber \\
&  & -\hbar\sum_{q=-1}^{+1}\sum_{s=1}^{3}\Omega_{s,q}^{\ast}\left(t\right)\exp\left(i\omega_{s}t\right)\left|-q\right\rangle \left\langle e\right|.\label{Hint(t)-1}
\end{eqnarray}
This is the interaction Hamiltonian whose effective version, for large
detunings to the red of the $\mathrm{D}_{2}$ line, allows us to realize
in the laboratory the GCDD Hamiltonian of Eq.~\eqref{Hlab-Upsilon}, in general up to an immaterial term proportional to the identity operator in the qutrit Hilbert space.
In the following we show how to do this through adiabatic elimination
of the excited state $|e\rangle $.

\subsection{Effective implementation of the GCDD Hamiltonian for the atomic qutrit}\label{Sec:EffectiveImplementation}

The unperturbed atomic Hamiltonian is written as
\begin{equation}
H_{\mathrm{atom}}  =  \hbar\omega_{g}\sum_{m=-1}^{1}\left|m\right\rangle \left\langle m\right|+\hbar\omega_{e}\left|e\right\rangle \left\langle e\right|,\label{Hatom}
\end{equation}
where the energy $\hbar\omega_{g}$ of the ground states $|m\rangle$, for $m=-1,0,1$, and the energy $\hbar\omega_{e}$ of the excited state $|e\rangle$ are such that
$\hbar(\omega_{e}-\omega_{g})$
is equal to the energy corresponding to the $\mathrm{D}_{2}$ line,
with wavelength given by $780.241$ nm, which corresponds to a frequency
of the order of $384.230$ THz:
\begin{equation}
\frac{\omega_{e}-\omega_{g}}{2\pi} \approx  384.230\textrm{ THz}.\label{D2line}
\end{equation}
Using Eq.~\eqref{Hatom} as the unperturbed Hamiltonian in the usual
interaction picture with the interaction Hamiltonian of Eq.~\eqref{Hint(t)-1},
we have
\begin{eqnarray}
H_{I}\left(t\right) & = & \exp\left(i\frac{H_{\mathrm{atom}}}{\hbar}t\right)H_{\mathrm{int}}\left(t\right)\exp\left(-i\frac{H_{\mathrm{atom}}}{\hbar}t\right)\nonumber \\
& = & -\hbar\sum_{q=-1}^{+1}\sum_{s=1}^{3}\Omega_{s,q}\left(t\right)\exp\left(-i\Delta_{s}t\right)\left|e\right\rangle \left\langle -q\right| \nonumber \\
&&
-\hbar\sum_{q=-1}^{+1}\sum_{s=1}^{3}\Omega_{s,q}^{\ast}\left(t\right)\exp\left(i\Delta_{s}t\right)\left|-q\right\rangle \left\langle e\right|,\label{HI}
\end{eqnarray}
where we recall that the detunings are defined by
\begin{equation}
\Delta_{s}  \equiv  \omega_{s}-\omega_{e}+\omega_{g}.\label{Deltas}
\end{equation}
The coherence times involved in superpositions of atomic quantum
states are typically of the order of a second or longer \cite{Treulein,Deutsch,Raman}. 
Thus, because the quantum-gate operation $\tau$ is an integer multiple
of $t_{0}$ and should be shorter than these typical coherence times,
we can take, roughly,
\begin{equation}
t_{0}  \sim  0.1\textrm{ s}.\label{ordert0}
\end{equation}
Hence, because we take Eq.~\eqref{ordert0} as valid, we see that
\begin{eqnarray}
\frac{\omega_{0}}{2\pi} & = & \frac{1}{t_{0}}\sim10\textrm{ Hz}\label{orderomega0}
\end{eqnarray}
is the corresponding rough estimate we can take for $\omega_{0}$
[cf. Eq.~\eqref{omega0}]. As we show below, about ten Hz
for $\omega_{0}/{(2\pi)}$ are enough for the GCDD method to work. In other words,
about ten Hz corresponds to the order of magnitude of the Hamiltonians
we need to emulate the $H_{L}^{\prime}/(2\pi\hbar)$ and $H_{F}^{\prime}/(2\pi\hbar)$
operators of Eqs.~\eqref{HLprime} and \eqref{HFprime} [see also
Eqs.~\eqref{eq:HL}, \eqref{omega0}, and \eqref{eq:HF}].

To implement the GCDD method in the context of laser control of an
atomic qutrit, here we show how to use two-photon transitions.
We need detunings that are much greater in magnitude  than the typical
few MHz of the Rabi frequencies in modulus [cf. Eq.~\eqref{orderOmega}], so that
we can make the adiabatic elimination of the state $|e\rangle $ \cite{Raman}.
One can easily implement this, given the relatively large difference
in energies in the transitions indicated in Fig.~\ref{fig:D2line}.
As we can see in this figure, the detunings can be as large as a few
GHz, and still the states of the ground $F=2$ level do not get involved
in the transitions (they are at the energy corresponding to $6.8$ GHz above the $F=1$ states we use). Since the detunings we use are negative, meaning that the photons excite a virtual level well below
the $|e\rangle $ excited state, the higher excited states
are not going to interfere with our transition scheme. Moreover, the
$\mathrm{D}_{2}$-line natural line-width for $^{87}\mathrm{Rb}$
is of the order of $6$ MHz \cite{NIST}, so that laser photons detuned to the
red of the $\mathrm{D}_{2}$ transition frequency by a few GHz will
not practically populate the excited state $|e\rangle$. As we show
in detail below, the magnitudes involved in the effective two-photon
Hamiltonian are proportional to the square of Rabi frequencies divided
by the detuning, which can be substantially higher than the few tens
of MHz (at least) required for an efficient GCDD implementation. Typically,
if we use, roughly,
\begin{equation}
\frac{\left|\Delta_{s}\right|}{2\pi} \sim  1\textrm{ GHz},\label{abs-Delta_s}
\end{equation}
using Eq.~\eqref{orderOmega} we find that
\begin{equation}
	\frac{\left|\Omega_{s,m}\left(t\right)\right|^{2}}{2\pi\left|\Delta_{s}\right|}  \sim  1\textrm{ kHz}.\label{Omega-squared-over-abs-Delta_s}
\end{equation}
Hence, using large detunings as in Eq.~\eqref{abs-Delta_s}, we end
up with an effective Hamiltonian (explained below) that can have its
magnitude as in Eq.~\eqref{Omega-squared-over-abs-Delta_s},flexibly above
the minimal requirement of Eq.~\eqref{orderomega0} for the GCDD method
to work, as we have discussed above. We notice that the values for $|\Delta_{s}|$ associated to Eq.~\eqref{abs-Delta_s} are consistent with the rotating wave-approximation used to obtain Eq.~\eqref{Hint(t)} since  they are much smaller than the transition frequency $\omega_{e}-\omega_{g}$ and the laser frequencies $\omega_s$.

Now, let us write the interaction-picture state as
\begin{equation}
\left|\psi_{I}\left(t\right)\right\rangle   =  \sum_{m=-1}^{1}C_{m}\left(t\right)\left|m\right\rangle +C_{e}\left(t\right)\left|e\right\rangle ,\label{PsiI}
\end{equation}
since $|\psi_{I}(t)\rangle \in\mathscr{H}_{4}$.
We can introduce the following projection operators:
\begin{equation}
P_{g}  \equiv  \sum_{m=-1}^{1}\left|m\right\rangle \left\langle m\right|\label{Pg}
\end{equation}
and
\begin{equation}
P_{e}  \equiv  \left|e\right\rangle \left\langle e\right|.\label{Pe}
\end{equation}
We immediately see that
\begin{eqnarray}
\left|\psi_{I}\left(t\right)\right\rangle  & = & \left(P_{g}+P_{e}\right)\left|\psi_{I}\left(t\right)\right\rangle \nonumber \\
& = & P_{g}\left|\psi_{I}\left(t\right)\right\rangle +P_{e}\left|\psi_{I}\left(t\right)\right\rangle .\label{projected-PsiI}
\end{eqnarray}
From the interaction-picture Schr\"{o}dinger equation and Eq.~\eqref{projected-PsiI},
we obtain
\begin{eqnarray}
i\hbar\frac{d}{dt}\left|\psi_{I}\left(t\right)\right\rangle  & = & H_{I}\left(t\right)\left|\psi_{I}\left(t\right)\right\rangle \nonumber  \\
& = & H_{I}\left(t\right)P_{g}\left|\psi_{I}\left(t\right)\right\rangle +H_{I}\left(t\right)P_{e}\left|\psi_{I}\left(t\right)\right\rangle .\label{Schrodinger-PsiI}\nonumber \\
\end{eqnarray}
Therefore, by applying the projectors of, respectively, Eqs.~\eqref{Pg} and \eqref{Pe}, to both sides of Eq.~\eqref{Schrodinger-PsiI}, we obtain the coupled Schr\"{o}dinger equations:
\begin{eqnarray}
i\hbar\frac{d}{dt}P_{g}\left|\psi_{I}\left(t\right)\right\rangle   =  &&P_{g}H_{I}\left(t\right)P_{g}\left|\psi_{I}\left(t\right)\right\rangle  \nonumber \\
&& +P_{g}H_{I}\left(t\right)P_{e}\left|\psi_{I}\left(t\right)\right\rangle \label{PgPsiI}
\end{eqnarray}
and
\begin{eqnarray}
i\hbar\frac{d}{dt}P_{e}\left|\psi_{I}\left(t\right)\right\rangle   =   && P_{e}H_{I}\left(t\right)P_{g}\left|\psi_{I}\left(t\right)\right\rangle \nonumber \\
&& +P_{e}H_{I}\left(t\right)P_{e}\left|\psi_{I}\left(t\right)\right\rangle.\label{PePsiI}
\end{eqnarray}
From Eq.~\eqref{HI} it is evident that
\begin{equation}
P_{g}H_{I}\left(t\right)P_{g}  =  0 \quad
\mathrm{and } \quad
P_{e}H_{I}\left(t\right)P_{e}  =  0.
\end{equation}
Thus, Eqs.~\eqref{PgPsiI} and \eqref{PePsiI} become
\begin{equation}
i\hbar\frac{d}{dt}P_{g}\left|\psi_{I}\left(t\right)\right\rangle   =  P_{g}H_{I}\left(t\right)P_{e}\left|\psi_{I}\left(t\right)\right\rangle \label{PgPsiI-1}
\end{equation}
and
\begin{equation}
i\hbar\frac{d}{dt}P_{e}\left|\psi_{I}\left(t\right)\right\rangle   =  P_{e}H_{I}\left(t\right)P_{g}\left|\psi_{I}\left(t\right)\right\rangle .\label{PePsiI-1}
\end{equation}
By formally integrating Eq.~\eqref{PePsiI-1} we obtain
\begin{equation}
P_{e}\left|\psi_{I}\left(t\right)\right\rangle  =  P_{e}\left|\psi_{I}\left(0\right)\right\rangle -\frac{i}{\hbar}\int_{0}^{t}dt^{\prime}P_{e}H_{I}\left(t^{\prime}\right)P_{g}\left|\psi_{I}\left(t^{\prime}\right)\right\rangle .\label{PePsiI(t)}
\end{equation}

Our intention is to start with the atom in the ground-state subspace, that
is, the population of the excited state is initially zero. Thus, using
this fact, that is,
\begin{equation}
P_{e}\left|\psi_{I}\left(0\right)\right\rangle   =  0,
\end{equation}
in Eq.~\eqref{PePsiI(t)}, we obtain
\begin{eqnarray}
P_{e}\left|\psi_{I}\left(t\right)\right\rangle  & = & -\frac{i}{\hbar}\int_{0}^{t}dt^{\prime}P_{e}H_{I}\left(t^{\prime}\right)P_{g}\left|\psi_{I}\left(t^{\prime}\right)\right\rangle .\label{PePsiI(t)-1}
\end{eqnarray}
Substitution of Eq.~\eqref{PePsiI(t)-1} into Eq.~\eqref{PgPsiI-1}
gives
\begin{eqnarray}
&& i\hbar\frac{d}{dt}P_{g}\left|\psi_{I}\left(t\right)\right\rangle  \nonumber \\ & & = -\frac{i}{\hbar}P_{g}H_{I}\left(t\right)P_{e}\int_{0}^{t}dt^{\prime}P_{e}H_{I}\left(t^{\prime}\right)P_{g}\left|\psi_{I}\left(t^{\prime}\right)\right\rangle ,\label{effective-PgPsiI}
\end{eqnarray}
where we have used the fact that $P_{e}$ is a projector operator
and, therefore, $P_{e}^{2}=P_{e}$. From Eqs.~\eqref{HI}, \eqref{PsiI},
\eqref{Pg}, and \eqref{Pe}, we see that
\begin{equation}
P_{g}H_{I}\left(t\right)P_{e}  =  -\hbar\sum_{q=-1}^{+1}\sum_{s=1}^{3}\Omega_{s,q}^{\ast}\left(t\right)\exp\left(i\Delta_{s}t\right)\left|-q\right\rangle \left\langle e\right|\label{PgHIPe}
\end{equation}
and
\begin{eqnarray}
&&\int_{0}^{t}dt^{\prime}P_{e}H_{I}\left(t^{\prime}\right)P_{g}\left|\psi_{I}\left(t^{\prime}\right)\right\rangle  \nonumber \\
&& =  -\left|e\right\rangle \hbar\sum_{q=-1}^{+1}\sum_{s=1}^{3}\int_{0}^{t}dt^{\prime} \Omega_{s,q}\left(t^{\prime}\right) \exp\left(-i\Delta_{s}t^{\prime}\right)C_{-q}\left(t^{\prime}\right). \nonumber \\
\label{PeHIPg}
\end{eqnarray}
From Eqs.~\eqref{PsiI}, \eqref{Pg}, \eqref{effective-PgPsiI}, \eqref{PgHIPe},
and \eqref{PeHIPg} we obtain
\begin{equation}
\frac{d}{dt}C_{m}\left(t\right)  =  -\sum_{q^{\prime}=-1}^{+1}\int_{0}^{t}dt^{\prime}K_{m,-q^{\prime}}\left(t,t^{\prime}\right)C_{-q^{\prime}}\left(t^{\prime}\right),\label{dpsiI(t)dt}
\end{equation}
for $m=-1,0,1$, where we have defined the kernel function:
\begin{eqnarray}
K_{m,-q^{\prime}}\left(t,t^{\prime}\right) & \equiv & \sum_{s=1}^{3}\sum_{s^{\prime}=1}^{3}\exp\left(i\Delta_{s}t-i\Delta_{s^{\prime}}t^{\prime}\right)\nonumber \\
&&\times\Omega_{s,-m}^{\ast}\left(t\right)\Omega_{s^{\prime},q^{\prime}}\left(t^{\prime}\right).\label{kernel}
\end{eqnarray}
We can also arrange Eqs.~\eqref{dpsiI(t)dt} and \eqref{kernel} in matrix format:
\begin{equation}
\frac{d}{dt}C\left(t\right)  =  -\int_{0}^{t}dt^{\prime}K\left(t,t^{\prime}\right)C\left(t^{\prime}\right),\label{matrix}
\end{equation}
where we have defined
\begin{equation}
C\left(t\right)  \equiv  \left[\begin{array}{c}
C_{-1}\left(t\right)\\
C_{0}\left(t\right)\\
C_{1}\left(t\right)
\end{array}\right]\label{C(t)}
\end{equation}
and
\begin{widetext}
\begin{equation}
K\left(t,t^{\prime}\right) \equiv  \sum_{s=1}^{3}\sum_{s^{\prime}=1}^{3}\exp\left(i\Delta_{s}t-i\Delta_{s^{\prime}}t^{\prime}\right) \times\left[\begin{array}{ccc}
\Omega_{s,1}^{\ast}\left(t\right)\Omega_{s^{\prime},1}\left(t^{\prime}\right) & \Omega_{s,1}^{\ast}\left(t\right)\Omega_{s^{\prime},0}\left(t^{\prime}\right) & \Omega_{s,1}^{\ast}\left(t\right)\Omega_{s^{\prime},-1}\left(t^{\prime}\right)\\
\Omega_{s,0}^{\ast}\left(t\right)\Omega_{s^{\prime},1}\left(t^{\prime}\right) & \Omega_{s,0}^{\ast}\left(t\right)\Omega_{s^{\prime},0}\left(t^{\prime}\right) & \Omega_{s,0}^{\ast}\left(t\right)\Omega_{s^{\prime},-1}\left(t^{\prime}\right)\\
\Omega_{s,-1}^{\ast}\left(t\right)\Omega_{s^{\prime},1}\left(t^{\prime}\right) & \Omega_{s,-1}^{\ast}\left(t\right)\Omega_{s^{\prime},0}\left(t^{\prime}\right) & \Omega_{s,-1}^{\ast}\left(t\right)\Omega_{s^{\prime},-1}\left(t^{\prime}\right)
\end{array}\right].\label{K}
\end{equation}
Iteration of Eq.~\eqref{matrix} yields:
\begin{equation}
C\left(t^{\prime}\right)  =  C\left(t\right)-\int_{t}^{t^{\prime}}dt_{1}\int_{0}^{t_{1}}dt_{2}K\left(t_{1},t_{2}\right)C\left(t\right) +(-1)^{2}\int_{t}^{t^{\prime}}dt_{1}\int_{0}^{t_{1}}dt_{2}K\left(t_{1},t_{2}\right)\int_{t}^{t_{2}}dt_{3}\int_{0}^{t_{3}}dt_{4}K\left(t_{3},t_{4}\right)C\left(t\right) +\dots\label{iteration}
\end{equation}
Let us calculate a generic element of the first kernel integral in
Eq.~\eqref{iteration}:
\begin{eqnarray}
\int_{t}^{t^{\prime}}dt_{1}\int_{0}^{t_{1}}dt_{2}K_{m,-q^{\prime}}\left(t_{1},t_{2}\right) & = & \sum_{s=1}^{3}\sum_{s^{\prime}=1}^{3}\int_{t}^{t^{\prime}}dt_{1}\int_{0}^{t_{1}}dt_{2}\exp\left(i\Delta_{s}t_{1}-i\Delta_{s^{\prime}}t_{2}\right)\Omega_{s,-m}^{\ast}\left(t_{1}\right)\Omega_{s^{\prime},q^{\prime}}\left(t_{2}\right)\nonumber \\
& = & \sum_{s=1}^{3}\sum_{s^{\prime}=1}^{3}\int_{t}^{t^{\prime}}dt_{1}\exp\left(i\Delta_{s}t_{1}\right)\Omega_{s,-m}^{\ast}\left(t_{1}\right)\int_{0}^{t_{1}}dt_{2}\exp\left(-i\Delta_{s^{\prime}}t_{2}\right)\Omega_{s^{\prime},q^{\prime}}\left(t_{2}\right),\label{Kelement}
\end{eqnarray}
\end{widetext}
where we have used Eq.~\eqref{kernel}. If we initially focus on the
integral over $t_{2}$ in Eq.~\eqref{Kelement}, we must make an assumption
about the time dependence of $\Omega_{s^{\prime},q^{\prime}}(t_{2})$.
Our aim here is to use Eq.~\eqref{HI} to effectively emulate the
GCDD Hamiltonian of Eq.~\eqref{Hlab-Upsilon}. As we show in the following,
although we are going to assume our Rabi frequencies $\Omega_{s^{\prime},q^{\prime}}(t_{2})/2\pi$ with  magnitudes of a few MHz [see Eq.~\eqref{orderOmega}], its time dependence is to be modulated in such a way that the corresponding spectral density results to be centered at about $\omega_{0}/2\pi$,
whose value is here assumed to be of the order of 10 Hz [see Eq.~\eqref{orderomega0}], with a width much smaller than $|\Delta_s|$. Let $G_{s^{\prime},q^{\prime}}(\omega)$
be the Fourier transform of $\Omega_{s^{\prime},q^{\prime}}(t^{\prime})$, namely,
\begin{equation}
G_{s^{\prime},q^{\prime}}\left(\omega\right)  \equiv  \frac{1}{2\pi}\int_{-\infty}^{+\infty}d\tau\exp\left(i\omega\tau\right)\Omega_{s^{\prime},q^{\prime}}\left(\tau\right),\label{Gsq}
\end{equation}
whose inverse is given by
\begin{equation}
\Omega_{s^{\prime},q^{\prime}}\left(t_{2}\right)  \equiv  \int_{-\infty}^{+\infty}d\omega\exp\left(-i\omega t_{2}\right)G_{s^{\prime},q^{\prime}}\left(\omega\right).\label{inverse-of-Gsq}
\end{equation}
Using Eq.~\eqref{inverse-of-Gsq}, we obtain
\begin{eqnarray}
&&\int_{0}^{t_{1}}dt_{2}\exp\left(-i\Delta_{s^{\prime}}t_{2}\right)\Omega_{s^{\prime},q^{\prime}}\left(t_{2}\right) \nonumber \\ & &=  \int_{0}^{t_{1}}dt_{2}\exp\left(-i\Delta_{s^{\prime}}t_{2}\right)\int_{-\infty}^{+\infty}d\omega\exp\left(-i\omega t_{2}\right)G_{s^{\prime},q^{\prime}}\left(\omega\right)\nonumber \\
& &= \int_{-\infty}^{+\infty}d\omega G_{s^{\prime},q^{\prime}}\left(\omega\right)\int_{0}^{t_{1}}dt_{2}\exp\left[-i\left(\Delta_{s^{\prime}}+\omega\right)t_{2}\right],\label{integral}
\end{eqnarray}
where we have changed the order of the integrals. Whatever form $G_{s^{\prime},q^{\prime}}(\omega)$
might have, it is assumed to be
characterized by a central value
$\omega_{0}>0$
and  a width both much smaller than $|\Delta_{s}|$,
so that we can write
\begin{eqnarray}
\int_{0}^{t_{1}}dt_{2}\exp\left[-i\left(\Delta_{s^{\prime}}+\omega\right)t_{2}\right] & = & \frac{\exp\left[-i\left(\Delta_{s^{\prime}}+\omega\right)t_{1}\right]-1}{-i\left(\Delta_{s^{\prime}}+\omega\right)}\nonumber \\
& \approx &  \frac{\exp\left[-i\left(\Delta_{s^{\prime}}+\omega\right)t_{1}\right]-1}{-i\Delta_{s^{\prime}}}. \nonumber \\ \label{result}
\end{eqnarray}
\begin{widetext}
Substituting Eq.~\eqref{result} into Eq.~\eqref{integral} we obtain
\begin{eqnarray}
 \int_{0}^{t_{1}}dt_{2}\exp\left(-i\Delta_{s^{\prime}}t_{2}\right)\Omega_{s^{\prime},q^{\prime}}\left(t_{2}\right) & & \approx  i\int_{-\infty}^{+\infty}d\omega G_{s^{\prime},q^{\prime}}\left(\omega\right)\frac{\exp\left[-i\left(\Delta_{s^{\prime}}+\omega\right)t_{1}\right]-1}{\Delta_{s^{\prime}}}\nonumber \\
& &  = \frac{i\Omega_{s^{\prime},q^{\prime}}\left(t_{1}\right)}{\Delta_{s^{\prime}}}\exp\left(-i\Delta_{s^{\prime}}t_{1}\right)-\frac{i\Omega_{s^{\prime},q^{\prime}}\left(0\right)}{\Delta_{s^{\prime}}},\label{result-1}
\end{eqnarray}
where we have used Eq.~\eqref{inverse-of-Gsq}. With the result of Eq.~\eqref{result-1} we can now tackle Eq.~\eqref{Kelement}:
\begin{eqnarray}
\int_{t}^{t^{\prime}}dt_{1}\int_{0}^{t_{1}}dt_{2}K_{m,-q^{\prime}}\left(t_{1},t_{2}\right)= && \sum_{s=1}^{3}\sum_{s^{\prime}=1}^{3}\frac{i}{\Delta_{s^{\prime}}}  \int_{t}^{t^{\prime}}dt_{1}   \exp\left[i\left(\Delta_{s}-\Delta_{s^{\prime}}\right)t_{1}\right]\Omega_{s,-m}^{\ast}\left(t_{1}\right)\Omega_{s^{\prime},q^{\prime}}\left(t_{1}\right)
\nonumber \\
&  & -\sum_{s=1}^{3}\sum_{s^{\prime}=1}^{3} \frac{\Omega_{s^{\prime},q^{\prime}}\left(0\right)}{\Delta_{s^{\prime}}\Delta_{s}}\left[\Omega_{s,-m}^{\ast}\left(t^{\prime}\right)\exp\left(i\Delta_{s}t^{\prime}\right)-\Omega_{s,-m}^{\ast}\left(t\right)\exp\left(i\Delta_{s}t\right)\right].  \label{double-integral}
\end{eqnarray}
We already see that all the terms in the second double sum on the right-hand
side of Eq.~\eqref{double-integral} are of second order in the quotient
between the order of magnitude of the Rabi frequencies [see Eq.~\eqref{orderOmega}]
and the order of magnitude of the detunings. Let us define this order
of magnitude more rigorously by assuming that, for all $t\in[0,t_{0}]$,
$s=1,2,3$, and $q=-1, 0,1$, we define $\para$ as the maximum absolute
value of $\Omega_{s,q}(t)/\Delta_{s}$, that is,
\begin{equation}
\para  \equiv  \max\left\{\left|\frac{\Omega_{s,q}\left(t\right)}{\Delta_{s}}\right|\right\}_{t\in\left[0,t_{0}\right],\ s\in\{1,2,3\},\ q\in\left\{ -1,0,1\right\} }.\label{order}
\end{equation}
Using the rough estimates of Eqs.~\eqref{orderOmega} and \eqref{abs-Delta_s}
we see that $\para$ can be even less than $10^{-3}$. Using Eq.~\eqref{inverse-of-Gsq}, we can now calculate the following integral:
\begin{eqnarray}
&&\int_{t}^{t^{\prime}}dt_{1}\exp\left[i\left(\Delta_{s}-\Delta_{s^{\prime}}\right)t_{1}\right]\Omega_{s,-m}^{\ast}\left(t_{1}\right)\Omega_{s^{\prime},q^{\prime}}\left(t_{1}\right) \nonumber \\ & &=  \int_{-\infty}^{+\infty}d\omega_{2}\int_{-\infty}^{+\infty}d\omega_{1}G_{s,-m}^{\ast}\left(\omega_{2}\right)G_{s^{\prime},q^{\prime}}\left(\omega_{1}\right)\int_{t}^{t^{\prime}}dt_{1}\exp\left[i\left(\Delta_{s}-\Delta_{s^{\prime}}+\omega_{2}-\omega_{1}\right)t_{1}\right].\label{second-integral}
\end{eqnarray}
Here we have two situations: $s\neq s^{\prime}$ and $s=s^{\prime}$.
Hence, taking these two cases into account, we obtain
\begin{eqnarray}
 \int_{t}^{t^{\prime}}dt_{1}\exp\left[i\left(\Delta_{s}-\Delta_{s^{\prime}}+\omega_{2}-\omega_{1}\right)t_{1}\right]  =&& \delta_{s,s^{\prime}}\int_{t}^{t^{\prime}}dt_{1}\exp\left[i\left(\omega_{2}-\omega_{1}\right)t_{1}\right]+\left(1-\delta_{s,s^{\prime}}\right)\frac{\exp\left[i\left(\Delta_{s}-\Delta_{s^{\prime}}+\omega_{2}-\omega_{1}\right)t^{\prime}\right]}{i\left(\Delta_{s}-\Delta_{s^{\prime}}+\omega_{2}-\omega_{1}\right)}\nonumber \\
&  &  -\left(1-\delta_{s,s^{\prime}}\right)\frac{\exp\left[i\left(\Delta_{s}-\Delta_{s^{\prime}}+\omega_{2}-\omega_{1}\right)t\right]}{i\left(\Delta_{s}-\Delta_{s^{\prime}}+\omega_{2}-\omega_{1}\right)}.\label{two-cases}
\end{eqnarray}
We now substitute Eq.~\eqref{two-cases} back into Eq.~\eqref{second-integral}.
We have assumed that the functions $G_{s,-m}^{\ast}(\omega_{2})$
and $G_{s^{\prime},q^{\prime}}(\omega_{1})$ only contribute
in a frequency region around $\omega_0$ much smaller than $|\Delta_s|$.
Thus, if we choose the detunings such that, for $s\neq s^{\prime}$,
the absolute difference $|\Delta_{s}-\Delta_{s^{\prime}}|$
is of the same order of magnitude of the $\max\{|\Delta_{s}|\}_{s\in\{ 1,2,3\} }$,
that is,
\begin{equation}
\left|\Delta_{s}-\Delta_{s^{\prime}}\right|  \sim  {\max\left\{\left|\Delta_{s}\right|\right\}_{s\in\left\{ 1,2,3\right\} }},\label{absolute-difference}
\end{equation}
which we estimate as about a few GHz [see Eq.~\eqref{abs-Delta_s}],
then we can assume $\Delta_{s}-\Delta_{s^{\prime}}+\omega_{2}-\omega_{1}\approx\Delta_{s}-\Delta_{s^{\prime}}$
in the denominators of Eq.~\eqref{two-cases} when this equation is
substituted back into Eq.~\eqref{second-integral}, obtaining
\begin{eqnarray}
 \int_{t}^{t^{\prime}}&&dt_{1}\exp\left[i\left(\Delta_{s}-\Delta_{s^{\prime}}\right)t_{1}\right]\Omega_{s,-m}^{\ast}\left(t_{1}\right)\Omega_{s^{\prime},q^{\prime}}\left(t_{1}\right) =  \delta_{s,s^{\prime}}\int_{t}^{t^{\prime}}dt_{1}\Omega_{s,-m}^{\ast}\left(t_{1}\right)\Omega_{s^{\prime},q^{\prime}}\left(t_{1}\right)\nonumber \\
&  & +\left(1-\delta_{s,s^{\prime}}\right)\frac{\Omega_{s,-m}^{\ast}\left(t^{\prime}\right)\Omega_{s^{\prime},q^{\prime}}\left(t^{\prime}\right)\exp\left[i\left(\Delta_{s}-\Delta_{s^{\prime}}\right)t^{\prime}\right]}{i\left(\Delta_{s}-\Delta_{s^{\prime}}\right)} -\left(1-\delta_{s,s^{\prime}}\right)\frac{\Omega_{s,-m}^{\ast}\left(t\right)\Omega_{s^{\prime},q^{\prime}}\left(t\right)\exp\left[i\left(\Delta_{s}-\Delta_{s^{\prime}}\right)t\right]}{i\left(\Delta_{s}-\Delta_{s^{\prime}}\right)}.\quad
\label{second-integral-1}
\end{eqnarray}
After substituting Eq.~\eqref{second-integral-1} into Eq.~\eqref{double-integral}
we end up with
\begin{eqnarray}
&& \int_{t}^{t^{\prime}}dt_{1}\int_{0}^{t_{1}}dt_{2}K_{m,-q^{\prime}}\left(t_{1},t_{2}\right)\nonumber \\
&  &  =  \!\sum_{s=1}^{3}\frac{i}{\Delta_{s}}\int_{t}^{t^{\prime}}\!dt_{1}
 \Omega_{s,-m}^{\ast}\left(t_{1}\right)
\Omega_{s,q^{\prime}}\left(t_{1}\right)\nonumber +\!\sum_{s=1}^{3}\sum_{s^{\prime}=1}^{3} \left(1-\delta_{s,s^{\prime}}\right)
\Omega_{s,-m}^{\ast}\left(t^{\prime}\right)
 \Omega_{s^{\prime},q^{\prime}}\left(t^{\prime}\right)
\frac{\exp\left[i\left(\Delta_{s}-\Delta_{s^{\prime}}\right)t^{\prime}\right]}{\Delta_{s^{\prime}}\left(\Delta_{s}-\Delta_{s^{\prime}}\right)}-\!\sum_{s=1}^{3}\sum_{s^{\prime}=1}^{3}  \left(1-\delta_{s,s^{\prime}}\right)
\nonumber \\
&  &\quad\times\frac{\Omega_{s,-m}^{\ast}\left(t\right)\Omega_{s^{\prime},q^{\prime}}\left(t\right)\exp\left[i\left(\Delta_{s}-\Delta_{s^{\prime}}\right)t\right]}{\Delta_{s^{\prime}}\left(\Delta_{s}-\Delta_{s^{\prime}}\right)} -\sum_{s=1}^{3}\sum_{s^{\prime}=1}^{3}
\frac{\Omega_{s^{\prime},q^{\prime}}\left(0\right)\left[\Omega_{s,-m}^{\ast}\left(t^{\prime}\right)\exp\left(i\Delta_{s}t^{\prime}\right) - \Omega_{s,-m}^{\ast}\left(t\right)\exp\left(i\Delta_{s}t\right)\right]}{\Delta_{s^{\prime}}\Delta_{s}}  .
\label{double-integral-result}
\end{eqnarray}
\end{widetext}
We conclude, therefore, that defining the kernel matrix including
terms with $s\neq s^{\prime}$, as in Eq.~\eqref{K}, amounts to producing
contributions of second order in $\para$ that are negligible when
compared with the terms with $s=s^{\prime}$, which are of first order
in $\para$, as we can directly verify in Eq.~\eqref{double-integral-result}
[cf. Eqs.~\eqref{order} and \eqref{absolute-difference}]. Therefore,
in the present problem, we keep only the $s=s^{\prime}$ terms in
Eq.~\eqref{double-integral-result} and neglect any other terms of
second order in $\para$:
\begin{eqnarray}
&&\int_{t}^{t^{\prime}}dt_{1}\int_{0}^{t_{1}}dt_{2}K_{m,-q^{\prime}}\left(t_{1},t_{2}\right) \nonumber \\
& & \approx \sum_{s=1}^{3}\frac{i}{\Delta_{s}}\int_{t}^{t^{\prime}}dt_{1}\Omega_{s,-m}^{\ast}\left(t_{1}\right)\Omega_{s,q^{\prime}}\left(t_{1}\right).\label{final-double-integral}
\end{eqnarray}
Now we can differentiate Eq.~\eqref{final-double-integral} with respect
to $t$ and get
\begin{equation}
\int_{0}^{t}dt_{2}K_{m,-q^{\prime}}\left(t,t_{2}\right)  \approx  \sum_{s=1}^{3}\frac{i}{\Delta_{s}}\Omega_{s,-m}^{\ast}\left(t\right)\Omega_{s,q^{\prime}}\left(t\right).\label{new-H}
\end{equation}
We see from the above discussion and Eqs.~\eqref{matrix} and \eqref{iteration}
that a time-local approximation of Eq.~\eqref{dpsiI(t)dt} is of first
order in $\para$ [see Eq.~\eqref{order}], and, using Eq.~\eqref{new-H},
we can write it as
\begin{eqnarray}
&&\frac{d}{dt}C_{m}\left(t\right)  =  -\sum_{q^{\prime}=-1}^{+1}\int_{0}^{t}dt^{\prime}K_{m,-q^{\prime}}\left(t,t^{\prime}\right)C_{-q^{\prime}}\left(t\right) +\mathcal{O}\left(\para^{2}\right) \nonumber \\
&& =  -\sum_{s=1}^{3}\sum_{q^{\prime}=-1}^{+1}\frac{i}{\Delta_{s}}\Omega_{s,-m}^{\ast}\left(t\right)\Omega_{s,q^{\prime}}\left(t\right)C_{-q^{\prime}}
\left(t\right)+\mathcal{O}\left(\para^{2}\right).\nonumber \\
\label{effective-eq}
\end{eqnarray}
	
Equation~\eqref{effective-eq} can be arranged in a matrix representation:
\begin{equation}
i\hbar\frac{d}{dt}C\left(t\right)\approx\sum_{s=1}^{3}H_{I,s}\left(t\right)C\left(t\right),\label{matrix-schrodinger}
\end{equation}
where, for $s=1,2,3$, we define
\begin{widetext}
\begin{equation}
H_{I,s}\left(t\right)  \equiv  \frac{\hbar}{\Delta_{s}}\left[\begin{array}{ccc}
\Omega_{s,1}^{\ast}\left(t\right)\Omega_{s,1}\left(t\right) & \Omega_{s,1}^{\ast}\left(t\right)\Omega_{s,0}\left(t\right) & \Omega_{s,1}^{\ast}\left(t\right)\Omega_{s,-1}\left(t\right)\\
\Omega_{s,0}^{\ast}\left(t\right)\Omega_{s,1}\left(t\right) & \Omega_{s,0}^{\ast}\left(t\right)\Omega_{s,0}\left(t\right) & \Omega_{s,0}^{\ast}\left(t\right)\Omega_{s,-1}\left(t\right)\\
\Omega_{s,-1}^{\ast}\left(t\right)\Omega_{s,1}\left(t\right) & \Omega_{s,-1}^{\ast}\left(t\right)\Omega_{s,0}\left(t\right) & \Omega_{s,-1}^{\ast}\left(t\right)\Omega_{s,-1}\left(t\right)
\end{array}\right].\label{HIs}
\end{equation}
\end{widetext}
	
We see, in Eq.~\eqref{matrix-schrodinger}, that, effectively,
we have found a Hamiltonian in the interaction picture
given by $\sum_{s=1}^{3}H_{I,s}(t)$. Now,
\begin{eqnarray}
C\left(t\right) & = & \sum_{m=-1}^{1}C_{m}\left(t\right)\left|m\right\rangle  =  P_{g}\left|\psi_{I}\left(t\right)\right\rangle \nonumber \\
& = & P_{g}\exp\left(i\frac{H_{\mathrm{atom}}}{\hbar}t\right)\left|\psi_{S}\left(t\right)\right\rangle ,\label{explain01}
\end{eqnarray}
where $|\psi_{S}(t)\rangle $ is in
the Schr\"{o}dinger picture [see Eq.~\eqref{HI}  for the definition of the interaction picture in our problem]. Using
Eqs.~\eqref{Hatom} and \eqref{Pg}, we see that
\begin{equation}
C\left(t\right)  =  \exp\left(i\omega_{g}t\right)P_{g}\left|\psi_{S}\left(t\right)\right\rangle ,
\end{equation}
since in this case $P_{g}=\mathbb{I}_{3}$, the identity
operator acting on $\mathscr{H}_{3}$ [cf. Eq.~\eqref{Pg}].
The above equation simply means that once computed the evolution of the state  $C(t)$,
the corresponding state in Schr\"{o}dinger picture state is obtained by multiplying it for the immaterial global phase factor $\exp(-i\omega_{g}t)$.
It follows that the dynamics in the qutrit subspace $\mathscr{H}_{3}$ is governed by the effective Hamiltonian
\begin{equation}
H_{\mathrm{eff}}\left(t\right)  = \hbar \omega_g \mathbb{I}_{3}+ \sum_{s=1}^{3}H_{I,s}\left(t\right).\label{Heff0}
\end{equation}
	
To make this scheme work, we know that $\Delta_{s}\in\mathbb{R}$
and $\Delta_{s}<0$. Therefore, we adopt the convention that
\begin{equation}
\sqrt{\Delta_{s}}  =  i\sqrt{-\Delta_{s}},\label{convention}
\end{equation}
so that
\begin{equation}
\left(\sqrt{\Delta_{s}}\right)^{\ast}  =  -i\sqrt{-\Delta_{s}}=  -\sqrt{\Delta_{s}},\label{Delta-ast}
\end{equation}
where $\sqrt{-\Delta_{s}}\in\mathbb{R}$ and $\sqrt{-\Delta_{s}}>0$.
Thus, Eq.~\eqref{HIs} becomes
\begin{equation}
H_{I,s}\left(t\right)
=  -\hbar\left[\begin{array}{c}
\frac{\Omega_{s,1}^{\ast}\left(t\right)}{\left(\sqrt{\Delta_{s}}\right)^{\ast}}\\
\frac{\Omega_{s,0}^{\ast}\left(t\right)}{\left(\sqrt{\Delta_{s}}\right)^{\ast}}\\
\frac{\Omega_{s,-1}^{\ast}\left(t\right)}{\left(\sqrt{\Delta_{s}}\right)^{\ast}}
\end{array}\right]\left[\begin{array}{ccc}
\frac{\Omega_{s,1}\left(t\right)}{\sqrt{\Delta_{s}}} & \frac{\Omega_{s,0}\left(t\right)}{\sqrt{\Delta_{s}}} & \frac{\Omega_{s,-1}\left(t\right)}{\sqrt{\Delta_{s}}}\end{array}\right],\label{HIs-1}
\end{equation}
since, from Eq.~\eqref{Delta-ast}, it follows
that
\begin{equation}
\left(\sqrt{\Delta_{s}}\right)^{\ast}\sqrt{\Delta_{s}}  =  -\sqrt{\Delta_{s}}\sqrt{\Delta_{s}}=  -\Delta_{s}.
\label{sqrtDelta-ast-sqrtDelta}
\end{equation}
Based on Eqs.~\eqref{Heff0} and \eqref{HIs-1} we can now express
$H_{\mathrm{eff}}(t)$ in a way that is analogous to Eq.~\eqref{Hlab-Upsilon},
allowing us to connect the Rabi frequencies of this Appendix with the
elements of the operator $\Upsilon$ of Sec.~\ref{Sec:The laboratory Hamiltonian}:
\begin{eqnarray}
H_{\mathrm{eff}}\left(t\right) & = & \hbar \omega_g \mathbb{I}_{3}\nonumber \\
&& -\hbar\sum_{s=1}^{3}\left[\begin{array}{c}
\frac{\Omega_{s,1}^{\ast}\left(t\right)}{\left(\sqrt{\Delta_{s}}\right)^{\ast}}\\
\frac{\Omega_{s,0}^{\ast}\left(t\right)}{\left(\sqrt{\Delta_{s}}\right)^{\ast}}\\
\frac{\Omega_{s,-1}^{\ast}\left(t\right)}{\left(\sqrt{\Delta_{s}}\right)^{\ast}}
\end{array}\right]\left[\begin{array}{ccc}
\frac{\Omega_{s,1}\left(t\right)}{\sqrt{\Delta_{s}}} & \frac{\Omega_{s,0}\left(t\right)}{\sqrt{\Delta_{s}}} & \frac{\Omega_{s,-1}\left(t\right)}{\sqrt{\Delta_{s}}}\end{array}\right]\nonumber \\
& = & \hbar \omega_g \mathbb{I}_{3} -\hbar\Theta^{\dagger}\left(t\right)\Theta\left(t\right),\label{OmegadaggerOmega}
\end{eqnarray}
where we have defined
\begin{equation}
\Theta\left(t\right)  \equiv  \left[\begin{array}{ccc}
\frac{\Omega_{1,1}\left(t\right)}{\sqrt{\Delta_{1}}} & \frac{\Omega_{1,0}\left(t\right)}{\sqrt{\Delta_{1}}} & \frac{\Omega_{1,-1}\left(t\right)}{\sqrt{\Delta_{1}}}\\
\frac{\Omega_{2,1}\left(t\right)}{\sqrt{\Delta_{2}}} & \frac{\Omega_{2,0}\left(t\right)}{\sqrt{\Delta_{2}}} & \frac{\Omega_{2,-1}\left(t\right)}{\sqrt{\Delta_{2}}}\\
\frac{\Omega_{3,1}\left(t\right)}{\sqrt{\Delta_{3}}} & \frac{\Omega_{3,0}\left(t\right)}{\sqrt{\Delta_{3}}} & \frac{\Omega_{3,-1}\left(t\right)}{\sqrt{\Delta_{3}}}
\end{array}\right].\label{Theta}
\end{equation}
Then, if we choose the Rabi frequencies and detunings
appearing in Eq.~\eqref{Theta} so that $\Theta(t)$ is Hermitian, we
can identify it with $\Upsilon(t)$ of Eq.~\eqref{Hlab-Upsilon} and this is how we
can implement, up to an immaterial term proportional to $\mathbb{I}_{3}$, the GCDD method for an atomic qutrit manipulated using
two-photon transitions. We notice that the immaterial terms proportional to $\mathbb{I}_{3}$ in Eqs.~\eqref{Hlab-Upsilon} and \eqref{OmegadaggerOmega} can be made equal if the atomic energy scale can be modified in such a way that $\omega_g=\omega_l$ in the new scale  [see Eq.~\eqref{omegag} for the definition of $\omega_l$]. Accordingly, thus, we choose, along the diagonal
of Eq.~\eqref{Theta}:
\begin{eqnarray}
\frac{\Omega_{1,1}\left(t\right)}{\sqrt{\Delta_{1}}} & = & \frac{\Omega_{1,1}^{\ast}\left(t\right)}{\left(\sqrt{\Delta_{1}}\right)^{\ast}},
\nonumber \\
\frac{\Omega_{2,0}\left(t\right)}{\sqrt{\Delta_{2}}} & = & \frac{\Omega_{2,0}^{\ast}\left(t\right)}{\left(\sqrt{\Delta_{2}}\right)^{\ast}},
\nonumber \\
\frac{\Omega_{3,-1}\left(t\right)}{\sqrt{\Delta_{3}}} & = & \frac{\Omega_{3,-1}^{\ast}\left(t\right)}{\left(\sqrt{\Delta_{3}}\right)^{\ast}}.
\end{eqnarray}
Using Eqs.~\eqref{convention} and \eqref{Delta-ast}, we then obtain:
\begin{eqnarray}
\Omega_{1,1}^{\ast}\left(t\right) & = & -\Omega_{1,1}\left(t\right),\nonumber \\
\Omega_{2,0}^{\ast}\left(t\right) & = & -\Omega_{2,0}\left(t\right), \nonumber \\
\Omega_{3,-1}^{\ast}\left(t\right) & = & -\Omega_{3,-1}\left(t\right).\label{omegadiag}
\end{eqnarray}
For the off-diagonal elements of Eq.~\eqref{Theta}, we choose:
\begin{eqnarray}
\frac{\Omega_{2,1}\left(t\right)}{\sqrt{\Delta_{2}}} & = & \frac{\Omega_{1,0}^{\ast}\left(t\right)}{\left(\sqrt{\Delta_{1}}\right)^{\ast}},\nonumber \\
\frac{\Omega_{3,1}\left(t\right)}{\sqrt{\Delta_{3}}} & = & \frac{\Omega_{1,-1}^{\ast}\left(t\right)}{\left(\sqrt{\Delta_{1}}\right)^{\ast}},\nonumber \\
\frac{\Omega_{3,0}\left(t\right)}{\sqrt{\Delta_{3}}} & = & \frac{\Omega_{2,-1}^{\ast}\left(t\right)}{\left(\sqrt{\Delta_{2}}\right)^{\ast}}.\label{omeganondiag}
\end{eqnarray}
Now, we are able to identify the remaining independent elements of
$\Theta(t)$ with those of $\Upsilon(t)$. By
imposing Eqs.~\eqref{omegadiag} and \eqref{omeganondiag}, and that $\Theta(t)=\Upsilon(t)$,
we obtain:
\begin{eqnarray}
\Omega_{1,1}\left(t\right) & = & \sqrt{\Delta_{1}}\Upsilon_{0,0}\left(t\right),\nonumber \\
\Omega_{1,0}\left(t\right) & = & \sqrt{\Delta_{1}}\Upsilon_{0,1}\left(t\right),\nonumber \\
\Omega_{1,-1}\left(t\right) & = & \sqrt{\Delta_{1}}\Upsilon_{0,2}\left(t\right),\nonumber \\
\Omega_{2,0}\left(t\right) & = & \sqrt{\Delta_{2}}\Upsilon_{1,1}\left(t\right),\nonumber \\
\Omega_{2,-1}\left(t\right) & = & \sqrt{\Delta_{2}}\Upsilon_{1,2}\left(t\right),\nonumber \\
\Omega_{3,-1}\left(t\right) & = & \sqrt{\Delta_{3}}\Upsilon_{2,2}\left(t\right).\label{OmegaUpsilon}
\end{eqnarray}
	
We observe that the above derivation leading to the effective Hamiltonian of Eq.~\eqref{OmegadaggerOmega}  could be coherently obtained also for values of $t_0$ different from the one choosen in Eq.~\eqref{ordert0} but satisfying the conditions required for performing the various approximations involved in the derivation. In this sense, we do not fix a specific value for $t_0$ in the numerical simulations based on Appendix~\ref{Sec:noise} and depicted in Fig.~\ref{fig:protection}. Consequently, the value of the gate time $\tau$ is not specified in these simulations and the other quantities are given in units of it.

\section{Environmental noise due to bosonic thermal baths} \label{Sec:noise}

Here, we explain how we use two baths of thermal bosons to simulate the
perturbations caused by the noisy environment considered in Sec.~\ref{numsim} (see also some previous
works of some of us on continuous dynamical decoupling of qubit systems \cite{Fanchini2007a, Fanchini2007b, Fanchini2007c, Fanchini2015}).

In the picture obtained by unitarily transforming the total Hamiltonian of the system and the environment of Eq.~\eqref{eq:Htotal} using $U_{c}(t)$, we obtain the total Hamiltonian in the control picture given in
Eq.~\eqref{eq:H}:
\begin{eqnarray}
H(t) & = & H_{G}\otimes\mathbb{I}_{E}+\mathbb{I}_{d}\otimes H_{E}+\left[U_{c}^{\dagger}\left(t\right)\otimes\mathbb{I}_{E}\right]H_{\mathrm{int}}\nonumber \\
&&\times\left[U_{c}\left(t\right)\otimes\mathbb{I}_{E}\right].
\end{eqnarray}
As explained in Sec.~\ref{numsim}, in our example we consider a qutrit subject to independent amplitude damping and dephasing noises. We divide the interaction Hamiltonian and the
environment free Hamiltonian in two parts, i.e., $H_{\mathrm{int}}=H_{\mathrm{int}}^{(1)}+H_{\mathrm{int}}^{(2)}$
and $H_{E}=H_{E}^{(1)}+H_{E}^{(2)}$, where the superscripts $1$
and $2$ refer, respectively, to the amplitude damping bosonic bath and to the dephasing one. Here,
we suppose that the two baths are identical, besides being independent. The above Hamiltonian terms are given in terms of the usual lowering and raising operators $a_{k}^{(i)}$ and $a_{k}^{(i)\dagger}$ for each mode $k$ of the $i$-$th$ bath, with $i=1,2$. In particular, the
 first interaction-Hamiltonian term
(which introduces the damping noise) is given by
\begin{eqnarray}
H_{\mathrm{int}}^{(1)} & = &  \lambda_{0,-1}^{(1)}\left(|0\rangle\langle -1| \otimes B^{(1)}+|-1\rangle\langle 0| \otimes B^{(1)\dagger}\right) \nonumber\\
 & &+  \lambda_{0,1}^{(1)}\left(|0\rangle\langle 1|\otimes B^{(1)}+|1\rangle\langle 0| \otimes B^{(1)\dagger}\right),\nonumber\\
 & = & \Lambda^{(1)} \otimes B^{(1)}+\Lambda^{(1)\dagger} \otimes B^{(1)\dagger},
\end{eqnarray}
where $B^{(1)}=   \sum_{k}  \hbar g_{k}a_{k}^{(1)}$ and $\Lambda^{(1)}=\lambda_{0,-1}^{(1)}(|0\rangle\langle -1|)+\lambda_{0,1}^{(1)}(|0\rangle\langle 1|)$, while the bath Hamiltonian associated to this class of error is given
by $H_{E}^{(1)}=\sum_{k} \hbar \omega_{k}a_{k}^{(1)\dagger}a_{k}^{(1)}$.
In a similar way, the second interaction-Hamiltonian term (which introduces
the dephasing noise) is given by
\begin{eqnarray}
H_{\mathrm{int}}^{(2)} & = &  \lambda_{0,-1}^{(2)}\left(|-1\rangle\langle -1|-|0\rangle\langle 0|\right)\otimes \left(B^{(2)}+B^{(2)\dagger}\right)\nonumber\\
 & & +  \lambda_{0,1}^{(2)}\left(|1\rangle\langle 1|-|0\rangle\langle 0|\right)\otimes \left(B^{(2)}+B^{(2)\dagger}\right),\nonumber\\
 & \equiv & \Lambda^{(2)}\otimes B^{(2)}+\Lambda^{(2)\dagger} \otimes B^{(2)\dagger},
\end{eqnarray}
where $B^{(2)}=\sum_{k} \hbar g_{k}a_{k}^{(2)}$ {and} $\Lambda^{(2)}=\lambda_{0,-1}^{(2)} (|-1\rangle\langle -1|-|0\rangle\langle 0|)+\lambda_{0,1}^{(2)}(|1\rangle\langle 1|-|0\rangle\langle 0|)$,
{while} the bath Hamiltonian associated to this class of error is given by
$H_{E}^{(2)}=\sum_{k}\hbar \omega_{k}a_{k}^{(2)\dagger}a_{k}^{(2)}$.

To obtain a master equation governing the three-level system dynamics,
we transform the total Hamiltonian to the interaction picture.
It is written as
\begin{equation}
\tilde{H}_{I}\left(t\right)=  \tilde{H}_{I}^{(1)}\left(t\right)+\tilde{H}_{I}^{(2)}\left(t\right),
\end{equation}
where
\begin{equation}
\tilde{H}_{I}^{(s)}\left(t\right)  =  \tilde{\Lambda}^{(s)}(t)\otimes\tilde{B}^{(s)}(t)+\tilde{\Lambda}^{(s)\dagger}(t)\otimes\tilde{B}^{(s)\dagger}(t),
\end{equation}
for $s=1,2$, with $\tilde{B}^{(s)}(t)=U_{E}^{(s)\dagger}(t)B^{(s)}U_{E}^{(s)}(t)$
and $\tilde{\Lambda}^{(s)}(t)=U_{G}^{\dagger}(t)U_{c}^{\dagger}(t)\Lambda^{(s)}U_{c}(t)U_{G}(t)$,
where $U_{G}(t)=\exp(-iH_{G}t/\hbar)$ and $U_{E}^{(s)}(t)=\exp(-iH_{E}^{(s)}t/\hbar)$.
With this transformation, the Redfield master equation is written as
\cite{Breuer}
\begin{eqnarray}
& & \frac{d\tilde{\rho}_{S}\left(t\right)}{dt}  = -\frac{1}{\hbar^2}\sum_{s=1}^{2}\int_{0}^{t}\mathrm{Tr}_{E}  \nonumber \\
&&\quad\left\{ \left[\tilde{H}_{I}^{(s)}\left(t\right), \left[\tilde{H}_{I}^{(s)}\left(t^{\prime}\right),\rho_{E}\otimes\tilde{\rho}_{S}\left(t\right)\right]\right]\right\} dt^{\prime},
\end{eqnarray}
where $\tilde{\rho}_{S}(t)=U_{G}^{\dagger}(t)U_{c}^{\dagger}(t)\rho_{S}(t)U_{c}(t)U_{G}(t)$
and $\rho_{E}$ is the environment density matrix, here given by a thermal
state, that is $\rho_{E}=\frac{1}{Z}\exp(-\beta\sum_{s=1}^{2}H_{{E}}^{(s)})$, where
$Z$ is the partition function $Z={\rm Tr}_{E}\{\exp(-\beta\sum_{s=1}^{2}H_{E}^{(s)})\}$, $\beta=1/(k_{B}T)$, $k_{B}$ is the Boltzmann constant, and $T$
is the temperature of the baths. We remark that the above Redfield master equation is derived under the Born approximation, linked to the assumption of weak coupling between the system and the environment, and to a part of the global Markovian approximation, so that the master equation is not Markovian \cite{Breuer}.

Now, substituting $\tilde{H}_{I}(t)$ into the master equation, we
finally obtain
\begin{eqnarray}
\frac{d\tilde{\rho}_{S}\left(t\right)}{dt}  =& -&\sum_{s=1}^{2}\int_{0}^{t}  \Big\{\left[\tilde{\rho}_{S}\left(t\right)\Lambda^{(s)\dagger}\left(t^{\prime}\right),\Lambda^{(s)}\left(t\right)\right]{\cal G}_{1}\left(t,t^{\prime}\right) \:\:\:\nonumber
\\
& -&\left[\Lambda^{(s)}\left(t^{\prime}\right)\tilde{\rho}_{S}\left(t\right),\Lambda^{(s)\dagger}\left(t\right)\right]{\cal G}_{1}^{\ast}\left(t,t^{\prime}\right)\nonumber\\
&  + & \left[\tilde{\rho}_{S}\left(t\right)\Lambda^{(s)}\left(t^{\prime}\right),\Lambda^{(s)\dagger}\left(t\right)\right]{\cal G}_{2}\left(t,t^{\prime}\right)\nonumber
\\
& -&\left[\Lambda^{(s)\dagger}\left(t^{\prime}\right)\tilde{\rho}_{S}\left(t\right),\Lambda^{(s)}\left(t\right)\right]{\cal G}_{2}^{\ast}\left(t,t^{\prime}\right)dt^{\prime}\Big\},
\end{eqnarray}
where the correlation functions ${\cal G}_{1}(t,t^{\prime})$ and ${\cal G}_{2}(t,t^{\prime})$ are given by
\begin{eqnarray}
{\cal G}_{1}\left(t,t^{\prime}\right) & = &\frac{1}{\hbar^2}\mathrm{Tr}_{E}\left\{ \tilde{B}^{(s)}(t)\rho_{E}\tilde{B}^{(s)\dagger}\left(t^{\prime}\right)\right\} ,\nonumber\\
{\cal G}_{2}\left(t,t^{\prime}\right) & = &\frac{1}{\hbar^2} \mathrm{Tr}_{E}\left\{ \tilde{B}^{(s)\dagger}\left(t\right)\rho_{E}\tilde{B}^{(s)}\left(t^{\prime}\right)\right\} .%
\end{eqnarray}
It is important to emphasize that, since we suppose identical baths,
the correlation functions are equivalent for $\tilde{B}^{(1)}$ and
$\tilde{B}^{(2)}$. Thus, the expressions for ${\cal G}_{1}(t,t^{\prime})$
and ${\cal G}_{2}(t,t^{\prime})$ are given by
\begin{eqnarray}
{\cal G}_{1}\left(t,t^{\prime}\right) & = & \sum_{k} |g_{k}|^{2}n_{k}\exp\left[-i\omega_{k}(t-t^{\prime})\right] ,\nonumber\\
{\cal G}_{2}\left(t,t^{\prime}\right) & = & \sum_{k}|g_{k}|^{2}\left(1+n_{k}\right)\exp\left[i\omega_{k}(t-t^{\prime})\right],
\end{eqnarray}
where $n_{k}=1/[\exp(\beta\hbar\omega_{k})-1]$ is the average number of photons in a mode with frequency $\omega_k$.
Finally, in the continuum limit, the sums become integrals and we obtain, using $s=t-t'$,
\begin{eqnarray}
\mathcal{G}_{1}\left(t,t^{\prime}\right) & = & \int_{0}^{\infty}d\omega J(\omega)n(\omega)\exp\left[-i\omega \left(t-t^{\prime}\right)\right],\qquad\qquad\nonumber\\
\mathcal{G}_{2}\left(t,t^{\prime}\right) & = & \int_{0}^{\infty}d\omega J(\omega) \left[1+n(\omega)\right] \exp\left[i\omega \left(t-t^{\prime}\right)\right],
\end{eqnarray}
where we have exploited the fact that the correlation functions are homogeneous in time,  $n(\omega)$ is the continuous frequency version of $n_{k}$,
namely, $n(\omega)=1/[\exp(\beta \hbar\omega)-1]$, and $J(\omega)$ is the spectral density. 

In the numerical simulations of Fig.~\ref{fig:protection}, we apply the GCDD to the qutrit considered in Sec.~\ref{Sec: atomic qutrit} and we choose for the spectral density $J(\omega)=\alpha^2 \ \omega\exp(-\omega/\omega_{c})$, where $\alpha$ is a dimensionless constant prefactor and $\omega_{c}$
is the angular cut-off frequency. We have also set equal all the $\lambda$ coupling constants, introducing as effective coupling parameter $\bar{\lambda} =\alpha \lambda_{0,-1}^{(s)}= \alpha \lambda_{0,1}^{(s)}=0.1$  for $ s=1, 2$,  and chosen  $\omega_c=4 \ \omega_{\mathrm{gate}} $, being $\omega_{\mathrm{gate}}=2 \pi/\tau$ where  $\tau$ is the gate time, and $\hbar \omega_c/(k_B T)=1$. As explained in Sec.~\ref{numsim}, we do not consider a specific value for the gate time $\tau$. The other quantities are then given in units of it.

\section{Extension to several qudits}\label{SeveralQudits}

Here, we explicitly show that the GCDD procedure presented in Secs.~\ref{Sec: The GCDD procedure} and \ref{Sec: Our prescription} for the case of a single qudit, can be extended to the case of an arbitrary number of qudits, identical or not. In particular, we consider the case where an arbitrary multi-qudit gate can act on the system which is also perturbed by the interaction with an environment which can contain both local and non-local terms. We do that by considering, for simplicity of notation, the case of two qudits. The extension to an ensemble of more than two qudits is  straightforward.

We consider two arbitrary qudits whose finite-dimensional Hilbert spaces  may be not isomorphic, that is, they have  dimensions $d_{1}$ and $d_{2}$, both integers, which may be different. Without loss of generality, we assume $ d_{2} \geqslant d_{1} > 1$. As for the one-qudit case, the free Hamiltonians of the qudits can be included in $H_G$ or assumed to be eliminated before the GCDD procedure.

The starting point is then a generalization of Eq.~\eqref{eq:Htotal}, which reads
\begin{equation}
H_{\mathrm{tot}}\left(t\right) =  \left[H_{c}\left(t\right) + H_{\mathrm{gate}}\left(t\right)\right]\otimes\mathbb{I}_{E}+\mathbb{I}_{d_1d_2}\otimes H_{E}+H_{\mathrm{int}},
\end{equation}
where, now, $H_{\mathrm{gate}}(t)$ corresponds to the action of a two-qudit
gate $H_G$, $\mathbb{I}_{d_1d_2}\equiv \mathbb{I}_{d_1}\otimes \mathbb{I}_{d_2}$, and the control Hamiltonian comprises only local one-qudit operations, that
is, $H_{c}(t)=H_c^{(1)}(t)\otimes \mathbb{I}_{d_2}+\mathbb{I}_{d_1}\otimes H_c^{(2)}(t)$,
where $H_c^{(1)}(t)$ and $H_c^{(2)}(t)$
in general are different, as described below. As for the noise, we consider
a generalized interaction Hamiltonian as:
\begin{equation}\label{eq:generalnoise}
	H_{\mathrm{int}}  =  \sum_{p=0}^{d_{1}-1}\sum_{q=0}^{d_{1}-1}\sum_{r=0}^{d_{2}-1}\sum_{s=0}^{d_{2}-1}
	\left|p\right\rangle \left\langle q\right|\otimes \left|r\right\rangle \left\langle s\right|\otimes B_{p,q,r,s},
\end{equation}
where $B_{p,q,r,s}$ are operators that act on the environmental states.
Let us notice that this Hamiltonian may contain, in general, both one-qudit perturbations (local noise)
as well as two-qudit perturbations (non-local noise). Then, we have $B_{p,q,r,s}=C_{p,q}\delta_{r,s}+D_{r,s}\delta_{p,q}+E_{p,q,r,s},$
so that
\begin{eqnarray}\label{eq:generalHint}
	H_{\mathrm{int}} & = & \sum_{p=0}^{d_{1}-1}\sum_{q=0}^{d_{1}-1}\left|p\right\rangle \left\langle q\right|\otimes\mathbb{I}_{d_{2}}\otimes C_{p,q}\nonumber\\
	&  &
	+\sum_{r=0}^{d_{2}-1}\sum_{s=0}^{d_{2}-1}\mathbb{I}_{d_{1}}\otimes \left|r\right\rangle \left\langle s\right| \otimes D_{r,s}\nonumber\\
	&  & +\sum_{p=0}^{d_{1}-1}\sum_{q=0}^{d_{1}-1}\sum_{r=0}^{d_{2}-1}\sum_{s=0}^{d_{2}-1}\left|p\right\rangle \left\langle q\right|\otimes \left|r\right\rangle \left\langle s\right|\otimes E_{p,q,r,s},\nonumber \\
\end{eqnarray}
where the terms involving the operators  $C_{p,q}$ and $D_{r,s}$ describe local noises, while the terms  involving the operators $E_{p,q,r,s}$ describe non-local noise.
As for the case of one qudit treated in Sec.~\ref{Sec: The GCDD procedure}, the condition there expressed in Eq.~\eqref{eq:dd} is not going
to be satisfied in our procedure, since the integral there appearing is going to be proportional to the
identity of the two qudits, tensorially
multiplied by an operator that only acts on the environment. Analogously to what has been done in Appendix~\ref{Sec:Dynamical decoupling condition}, it is possible to show that 
this result is sufficient to dynamically decouple the qudits from
the noisy perturbations. In the end,
the final state is the same that, up to first order of perturbation,
one would obtain in the absence of interaction with the environment.
In other words, Eq.~\eqref{eq:dd}, applied to the two-qudit case treated here, is sufficient, but
not necessary, since if the integral is proportional
to the qudits-system identity, instead of zero, this is sufficient for guaranteeing dynamical decoupling. This is due to the fact that the integral
appearing in Eq.~\eqref{eq:dd} is present in the exponential in the Magnus expansion of the total propagator at first order, so that, if it is proportional to the system's
identity, dynamical decoupling is still obtained (see Appendix~\ref{Sec:Dynamical decoupling condition}). 

It is then enough to obtain for the integral of
Eq.~\eqref{eq:dd}  something proportional to the two-qudit identity. To this aim, we use control Hamiltonians $H_c^{(1)}(t)$ and $H_c^{(2)}(t)$ which are different with each other and
whose respective propagators, $U_c^{(1)}(t)$ and $U_c^{(2)}(t)$,
are defined analogously to the single-qudit propagator given by Eq.~\eqref{eq:Uc(t)} as
\begin{equation}
U_{c}^{(j)}\left(t\right) \equiv \exp\!\left(\!-i\ome^{(j)} t\right)\exp\!\left(\!-i\frac{H^{(j)}_{L}}{\hbar}t\right)\exp\!\left(\!-i\frac{H^{(j)}_{F}}{\hbar}t\right)\!,
\label{eq:Uci(t)SM}
\end{equation}
where $j=1, 2,$ runs over the two qudits,
\begin{equation}
	H_{L}^{(j)}\left|k_{j}\right\rangle   \equiv  k_{j}\hbar\omega^{(j)}_{d_{j}}\left|k_{j}\right\rangle ,\label{eq:HLSM}
\end{equation}
for $k_{j}=0,1,\dots,d_{j}-1$, 
\begin{equation}
	\omega^{(j)}_{d_{j}} \equiv  d_{j} \,\omega^{(j)}_{0},\quad  \omega^{(1)}_{0}\equiv \omega_{0}, \quad\omega^{(2)}_{0}=  d_{2}^{2}\omega_{0}^{\left(1\right)},\label{eq:omegadSM}
\end{equation}
\begin{equation}
	H^{(j)}_{F}\left|\psi_{n_{j}}\right\rangle   \equiv  n_{j}\hbar\omega^{(j)}_{0}\left|\psi_{n_{j}}\right\rangle ,\label{eq:HFSM}
\end{equation}
for $n_{j}=0,1,\dots,d_{j}-1$, and
\begin{equation}
	\ome^{(j)}  \equiv - \frac{\mathrm{Tr}\left\{H^{(j)}_{L}\right\}+\mathrm{Tr}\left\{H^{(j)}_{F}\right\}}{\hbar d_{j}}.
\end{equation}
The main point is that, differently from the one-qudit case where we have used  $\omega_{0}$, here we make use of two
different frequencies,  $\omega_{0}^{(1)}$ and $\omega_{0}^{(2)}$.
Then, we define the control propagator for the two qudits as
\begin{equation}\label{eq:twoquidstUc}
U_{c}\left(t\right)  =  U_c^{(1)}\left(t\right)\otimes U_c^{(2)}\left(t\right).
\end{equation}
In order to show that the integral of
Eq.~\eqref{eq:dd} in the case considered here of two qudits is proportional to the two-qudit identity, we consider the integral for a generic term in the decomposition of  $H_{\mathrm{int}}$  in Eq.~\eqref{eq:generalnoise} and we exploit the fact that  the frequencies we have chosen amount to different
time scales. Proceeding analogously to what has been done in Sec.~\ref{Sec:U_c},  we obtain in the two-qudit version of Eq.~\eqref{eq:main}
\begin{widetext}
\begin{eqnarray}\label{eq:factored}
	&&\int_{0}^{t_{0}}dt\,U_{c}^{\dagger}\left(t\right)\left(\left|p\right\rangle \left\langle q\right|\otimes \left|r\right\rangle \left\langle s\right|\right)U_{c}\left(t\right)\nonumber  \\ && =\int_{0}^{t_{0}}dt \left[U_c^{(1)\,\dagger}\left(t\right)\left|p\right\rangle \left\langle q\right|U_c^{(1)}\left(t\right)\right] \otimes\left[U_c^{(2)\,\dagger}\left(t\right)\left|r\right\rangle \left\langle s\right|U_c^{(2)}\left(t\right)\right]\nonumber\\
	&&=	\frac{1}{d_{1} d_{2}}\sum_{m=0}^{d_{1}-1}\sum_{n=0}^{d_{1}-1}\sum_{m^{\prime}=0}^{d_{2}-1}\sum_{n^{\prime}=0}^{d_{2}-1}\exp\left(-\frac{2\pi i}{d_{1}}pn\right)\exp\left(\frac{2\pi i}{d_{1}}qm\right)\nonumber \exp\left(-\frac{2\pi i}{d_{2}}rn^{\prime}\right)\exp\left(\frac{2\pi i}{d_{2}}sm^{\prime}\right)\left|\psi_{n}\right\rangle \left\langle \psi_{m}\right|\otimes\left|\psi_{n^{\prime}}\right\rangle \left\langle \psi_{m^{\prime}}\right|
\nonumber \\ &&
	\quad\times\int_{0}^{t_{0}}dt\,\exp\left[i\left(\left(n-m\right)+\left(p-q\right)d_{1}+\left(n^{\prime}-m^{\prime}\right)d_{2}^{2}
	+\left(r-s\right)d_{2}^{3}\right)\omega_{0}t\right]	=\frac{t_{0}}{d_{1}d_{2}}\delta_{p,q}\delta_{r,s}\mathbb{I}_{d_{1}}\otimes\mathbb{I}_{d_{2}},
\end{eqnarray}
where we have used 
\begin{equation}
\frac{1}{\sqrt{d}}\sum_{n=0}^{d-1} \exp\left(-\frac{2\pi i}{d}kn\right)\left|\psi_{n}\right\rangle    \frac{1}{d}\sum_{j=0}^{d-1}\sum_{n=0}^{d-1}\exp\left[\frac{2\pi i}{d}\left(j-k\right)n\right]\left|j\right\rangle \frac{1}{d}\sum_{j=0}^{d-1}\sum_{n=0}^{d-1}\left\{ \exp\left[\frac{2\pi i}{d}\left(j-k\right)\right]\right\} ^{n}\left|j\right\rangle  \left|k\right\rangle ,
\end{equation}
and applied the following reasoning. The time integration appearing in Eq.~\eqref{eq:factored} before the final result is not zero
if and only if
\begin{equation}
	\left(n-m\right)+\left(p-q\right)d_{1}+\left(n^{\prime}-m^{\prime}\right)d_{2}^{2}+\left(r-s\right)d_{2}^{3} =  0,
\end{equation}
where $d_{1},d_{2}\in\mathbb{Z}$
with $d_{2}\geqslant d_{1}>1$. Given the constraints on the various parameters, the l.h.s. of the above equation 
can be zero only if $m=n$, $q=p$, $m^{\prime}=n^{\prime}$, and
$s=r$. It follows from Eqs.~\eqref{eq:generalnoise} and \eqref{eq:factored} that the integral in Eq.~\eqref{eq:dd} is here proportional to the two-qudit identity:
\begin{equation}
\int_{0}^{t_{0}}dt\,\left[U_{c}^{\dagger}\left(t\right)\otimes\mathbb{I}_{E}\right]H_{\mathrm{int}}\left[U_{c}\left(t\right)\otimes\mathbb{I}_{E}\right] =  \frac{t_{0}}{d_{1}d_{2}}\mathbb{I}_{d_{1}} \otimes \mathbb{I}_{d_{2}} \otimes
	\sum_{p=0}^{d_{1}-1}\sum_{r=0}^{d_{2}-1}
 B_{p,p,r,r}.\label{eq:dd2qudit}
\end{equation}
\end{widetext}

We emphasize again that this result is enough to eliminate any noise
involved in Eq.~\eqref{eq:generalnoise}, even when the environment involves non-local terms acting on the qudits, and that this result is only possible, in general, if
we can really separate all the time scales of each control propagator,
so that the final results of Eq.~\eqref{eq:factored} can be given in terms of factored Kronecker deltas. 
The choice made in Eq.~\eqref{eq:omegadSM} is motivated by this aim. In the case of a single qudit, treated in Secs.~\ref{Sec: The GCDD procedure} and \ref{Sec: Our prescription}, we need only two sets of time scales,
determined, respectively, by $\{ k\omega_{0},\, k= 0,1,\dots,d-1\} $
and $\{ kd\omega_{0},\, k=0,1,\dots,d-1\} $. Now, having two qudits, we need four different
sets of time scales in order to have in Eq.~\eqref{eq:factored} the time integral giving rise to several factored Kronecker deltas. According to our choice, we have $\{ k\omega_{0},\, k=0, 1,\dots,d_{1}-1\} $
and $\{ kd_{1}\omega_{0},\, k=0, 1,\dots,d_{1}-1 \} $ for one of the
qudits, while we use $\{kd_{2}^{2}\omega_{0},\, k=0,1,\dots,d_{2}-1\} $
and $\{ kd_{2}^{3}\omega_{0},\, k=0,1,\dots,d_{2}-1\} $ for the other
one. We finally stress that when $d_{1}=d_{2}$ all that has been said above is still valid and we can, therefore, protect two identical qudits using the same procedure.

The problem simplifies in the case of only local noises, that is, when in Eq.~\eqref{eq:generalHint} the terms with $E_{p,q,r,s}$ are missing and then $H_{\mathrm{int}}$ reduces to
\begin{eqnarray}
	H_{\mathrm{int}} & = & \sum_{p=0}^{d_1-1}\sum_{q=0}^{d_1-1}\left|p\right\rangle \left\langle q\right|\otimes\mathbb{I}_{d_2}\otimes C_{p,q} \nonumber \\
	&& +\sum_{r=0}^{d_2-1}\sum_{s=0}^{d_2-1}\mathbb{I}_{d_1}\otimes\left|r\right\rangle \left\langle s\right| \otimes D_{r,s}.
\end{eqnarray}
In this case, we can define each $U_c^{(j)}(t)$ operator as in Eq.~\eqref{eq:Uci(t)SM}, but using in Eq.~\eqref{eq:omegadSM}
$\omega^{(2)}_{0}=\omega^{(1)}_{0}= \omega_{0}$ instead of $\omega^{(2)}_{0}=  d_{2}^{2}\omega_{0}^{\left(1\right)}= d_{2}^{2}\omega_{0}$.
Indeed, under this assumption and using Eq.~\eqref{eq:main}, the terms needed to compute Eq.~\eqref{eq:dd} give  
\begin{eqnarray}
&&	\int_{0}^{t_{0}}dt\, U_{c}^{\dagger}\left(t\right)\left(\left|p\right\rangle \left\langle q\right|\otimes\mathbb{I}_{d_2}\right)U_{c}\left(t\right) 
=  \int_{0}^{t_{0}}dt
\nonumber \\	&&\left[U_c^{(1)\,\dagger}\left(t\right)\left|p\right\rangle \left\langle q\right|U_c^{(1)}\left(t\right)\right]\otimes
\left[U_c^{(2)\,\dagger}\left(t\right)\mathbb{I}_{d_2}U_c^{(2)}\left(t\right)\right]  \nonumber\\
	& & = \frac{t_{0}}{d_1}\delta_{p,q}\mathbb{I}_{d_1} \otimes \mathbb{I}_{d_2},
\end{eqnarray}
and
\begin{eqnarray}
&&	\int_{0}^{t_{0}}dt\, U_{c}^{\dagger}\left(t\right)\left(
\mathbb{I}_{d_1}\otimes\left|r\right\rangle \left\langle s\right|
\right)U_{c}\left(t\right) 
=  \int_{0}^{t_{0}}dt \nonumber \\	& &\left[U_c^{(1)\,\dagger}\left(t\right)
\mathbb{I}_{d_1}
U_c^{(1)}\left(t\right)\right]\otimes
\left[U_c^{(2)\,\dagger}\left(t\right) \left|r\right\rangle \left\langle s\right|
U_c^{(2)}\left(t\right)\right]  \nonumber\\
& & = \frac{t_{0}}{d_2}\delta_{r,s} \mathbb{I}_{d_1} \otimes \mathbb{I}_{d_2}.
\end{eqnarray}
It follows that the integral in Eq.~\eqref{eq:dd}  is still proportional to the two-qudit identity:
\begin{eqnarray}
&&\int_{0}^{t_{0}}dt\,\left[U_{c}^{\dagger}\left(t\right)\otimes\mathbb{I}_{E}\right]H_{\mathrm{int}}\left[U_{c}\left(t\right)\otimes\mathbb{I}_{E}\right] \nonumber\\
&&=   t_{0} \mathbb{I}_{d_1} \otimes \mathbb{I}_{d_2} \otimes \left( \frac{1}{d_1}\sum_{p=0}^{d-1} C_{p,p}  + \frac{1}{d_2}\sum_{r=0}^{d-1} D_{r,r}\right).\quad\label{eq:dd2quditlocal}
\end{eqnarray}
We notice that in the above simplified  case with only local noises, $U_c^{(1)}(t)=U_c^{(2)}(t)$ in the case of identical qudits, since in this case $d_1=d_2$.

Now, we address the problem of the two-qudit gate, irrespective whether the qudits are identical or not. Let us start noticing that since the control Hamiltonians are local operators acting
on each qudit, they do not change any possible entanglement between
the two qudits. We proceed in an analogous way as we did for the
case of one qudit, that is, we use
\begin{equation}
	H_{\mathrm{gate}}\left(t\right) =  U_{c}\left(t\right)H_{G}U_{c}^{\dagger}\left(t\right),
\end{equation}
where now, of course, $H_{G}$ is the desired two-qudit
gate and $U_{c}(t)$ is the two-qudit control
propagator defined in Eq.~\eqref{eq:twoquidstUc}. It might be very difficult to be able to implement such
an orchestrated two-qudit Hamiltonian, but, once it is done, we have
a two-qudit gate protected against a very general kind of perturbing
environment by using $H_{\mathrm{lab}}(t) =  H_{c}(t)+	H_{\mathrm{gate}}(t)$, where $H_{c}(t)$ is given by
\begin{eqnarray}
&&H_{c}\left(t\right)  =  i\hbar\frac{dU_{c}\left(t\right)}{dt}U_{c}^{\dagger}\left(t\right)\nonumber \\
	& &=    \sum_{j=1}^2 \left[\hbar\ome^{(j)}\mathbb{I}_{d_1} \otimes \mathbb{I}_{d_2} +H^{(j)}_{L}+U^{(j)}_{L}\left(t\right)H^{(j)}_{F}U_{L}^{(j)\,\dagger}\left(t\right)\right], \nonumber \\
	\label{eq:Hc(t)twoqudit}
\end{eqnarray}
where
\begin{equation}
U_{L}^{(j)}\left(t\right)  \equiv  \exp\left(-i\frac{H^{(j)}_{L}}{\hbar}t\right).\label{eq:ULtwoquidit}
\end{equation} 

A simple and direct application of the procedure explained in this Appendix is represented by the case when one does not want to protect the action of a gate on the qudits but only preserve their state since, for instance, it is an entangled state. In this case, the state of the two qudits can be preserved from the action of the noise,
as a protected memory state, by simply considering the case $H_{G}=0$, that is by using $H_{\mathrm{lab}}(t) =  H_{c}(t)$, where $H_{c}(t)$ is given in Eq.~\eqref{eq:Hc(t)twoqudit}.

\end{document}